\documentclass[lettersize,journal]{IEEEtran}
\usepackage{caption}
\usepackage{amsfonts}
\usepackage{amsmath}  
\usepackage{amssymb}  
\usepackage{amsthm} 
\usepackage{algorithmic}
\usepackage{algorithm}
\usepackage{array}
\usepackage[caption=false,font=normalsize,labelfont=sf,textfont=sf]{subfig}
\usepackage{textcomp}
\usepackage{stfloats}
\usepackage{url}
\usepackage{verbatim}
\usepackage{graphicx}
\usepackage{xcolor}
\usepackage{cite}
\usepackage{booktabs}
\usepackage{multirow}
\usepackage{hyperref}
\usepackage{commath}
\usepackage{subcaption} 
\usepackage{empheq}
\usepackage{stmaryrd}
\usepackage{enumitem}

\usepackage{tikz}
\usetikzlibrary{decorations.pathreplacing, arrows.meta}
\theoremstyle{definition}

\theoremstyle{remarkstyle}
\newtheorem{remark}{Remark}

\input{mysymbol.sty}

\def\bbx{{\ensuremath{\boldsymbol{x}}}}
\def\bbs{{\ensuremath{\boldsymbol{s}}}}

\begin{document}

\title{Topological Kalman Filtering on Cell Complexes}

\author{
    Chengen Liu, 
    Rohan Money,
    Ting Gao,
    Mohammad Sabbaqi,
    Baltasar Beferull-Lozano,~and Elvin Isufi
    
    \thanks{
        Chengen Liu, Mohammad Sabbaqi, and Elvin Isufi are with the Faculty of Electrical Engineering, Mathematics and Computer Science, Department of Intelligent Systems, TU Delft (\{c.liu-15, M.Sabbaqi, e.isufi-1\}@tudelft.nl). Ting Gao is with Faculty of Civil Engineering and Geosciences, TU Delft (T.Gao-1@tudelft.nl). Baltasar Beferull-Lozano and Rohan T. Money are with the SIGIPRO Department, SIMULA Metropolitan, Norway. This paper is supported by the NWO Grant Veni (No. 222.032) financed by the Netherlands Organization for Scientific Research, supported by the SURE-AI Center grant 357482 and DISCO grant 338740, Research Council of Norway. Chengen Liu receives funding from the China Scholarship Council.
    }
}

\markboth{}%
{Shell \MakeLowercase{\textit{et al.}}: A Sample Article Using IEEEtran.cls for IEEE Journals}

\maketitle

\begin{abstract}
Inferring latent dynamics from multivariate time-series defined over topological cell complexes is crucial for capturing the complex, higher-order interactions inherent in real-world systems such as in water, sensor, and transportation networks. However, reconstructing these latent states is challenging because the signals are coupled across higher-order topologies, while high dimensionality, nonlinear observations, and unknown structures increase the difficulty. To address this, we propose a topology-aware state space framework derived from stochastic partial differential equations on cell complexes. State evolution follows heat-like topological diffusion, with perturbations propagating along boundary operators. Under partial observability, we model observations using a cell complex convolution of latent states coupled with a nonlinear mapping. We perform recursive state estimation via an Extended Kalman Filter, simultaneously learning model parameters and uncertainties through an online Expectation-Maximization algorithm. Finally, for scenarios where only lower-order topological structure is known, e.g., nodes and edges, as in critical infrastructure networks, we introduce a heuristic cell identification algorithm to explicitly infer the second-order cell structures. Validations on synthetic and real datasets from water, sensor and transportation networks demonstrate that our approach yields reliable estimates under partial observability and successfully recovers the underlying topological structures.
\end{abstract}


\section{Introduction}

\IEEEPARstart{S}{tate} estimation serves as the fundamental method for inferring latent system variables from noisy observations by coupling stochastic dynamics with measurement models~\cite{kalman1960new,course2023state}. Modern systems, such as neural pathways, traffic networks, and social networks, generate high dimensional time-series data, in which deriving closed-form analytical system and observation models is often intractable. However, these systems do not evolve randomly. Their behavior is heavily constrained by an underlying physical or abstract topology, such as synaptic wiring, road connections, and social contacts. In this work, we use this topological information as an inductive bias to build a topology-aware state space model and perform online state estimation from partial and noisy observations. This inductive bias constrains the system states to evolve along the topology of the cell complex. This structural prior prevents over-parameterization and makes the state space model amenable to low-data regimes.

While graphs have conventionally been employed to encode pairwise data relationships~\cite{grady2010discrete} and have been successfully incorporated into state-space models~\cite{sabbaqi2025gknet, buchnik2024gsp, tenorio2025tracking, elvira2022graphical, isufi2025topological, millan2025topology}, they are fundamentally restricted to dyadic interactions and signals residing exclusively on vertices. This limitation precludes graph-based models from capturing higher-order dependencies or processing dynamic signals defined over higher-dimensional topological spaces (e.g., edges, faces). Graphs fail to account for the cross-order coupling of signals that coexist across these distinct topological orders. Such complex dependencies are prevalent in physical systems like water distribution networks, where the latent state spans multiple topological orders: nodal pressures (vertices), pipe discharges (edges), and flow circulations  (higher-order faces). In these networks, signals are not independent but are rigorously constrained by multi-way physical  laws governed by the underlying topology.

To capture both these multi-way dependencies and the cross-order nature of the data, time-series can be modeled over higher-order topological domains, such as cell complexes, which generalize graphs by enabling the unified processing of signals residing simultaneously across nodes, edges, and faces~\cite{topelv2025,bick2023higher,salnikov2018simplicial, Rohan_SPL}. Within this framework, algebraic boundary operators serve as the fundamental mappings between adjacent topological orders. These operators explicitly define how, for example, edge flows diverge into nodal potentials or how face circulations project onto boundary edges~\cite{calmon2023dirac,baccini2022weighted}. Building upon these boundary operators, topological operators such as Hodge Laplacians and the Dirac operator provide a mathematically principled mechanism to govern signal diffusion and cross-order coupling~\cite{bick2023higher,salnikov2018simplicial}. Motivated by these structural properties, recent advances have established a comprehensive toolbox for analyzing data on topological spaces~\cite{isufi2024topological, barbarossa2020topological, schaub2021signal}. These foundational developments span topological convolutions and filtering~\cite{sardellitti2024topological, grimaldi2025topological, yang2022simplicial, yang2022simplicialtrend}, neural architectures~\cite{yang2022simplicialconvolutional, battiloro2024generalized, isufi2025topological}, spectral analyses~\cite{barbarossa2020topological}, signal recovery~\cite{reddy2024recovery, liu2023unrolling}, and topological detection and random walks~\cite{liu2025matched, schaub2020random}.

Building upon these foundational tools, there is a growing emphasis on time-series modeling over cell and simplicial complexes to capture, cross-order spatiotemporal dependencies \cite{krishnan2024simplicial, Rohan_SPL, toplor2025, impduc2025}.
Notable contributions to topological time-series modeling include simplicial vector autoregressive models, which leverage simplicial convolutions to capture spatial relationships \cite{krishnan2024simplicial} and combines the autoreggrsvie nature of vector autoregressive model. For streaming scenarios, a least mean squares (LMS) approach was recently introduced to learn time-series models from sequential data \cite{marinucci2025topological}. Additionally, a topology-aware state-space model has been proposed for processing multiple time-series and imputing  signals \cite{Rohan_SPL}.

Despite these advances, the majority of current topological models for time-series implicitly assume that the measured signals directly represent a fully observable system state. In practice, however, the dynamics of many real-world systems are governed by latent states that are not directly measured. For example, in water distribution networks, typically only a sparse set of flow and pressure measurements is available, whereas the complete hydraulic state—comprising nodal heads, pipe flows, and underlying demand patterns—remains latent \cite{sabbaqi2025gknet, kerimov2024towards, cattai2025physics}. Standard data-driven models often struggle under these conditions because they lack mechanisms to infer unobserved variables. The state-space model in \cite{Rohan_SPL} provides a valuable approach for handling missing data on topological structures; however, it utilizes a predetermined, binary mask for its observation matrix. This implies that the observations are modeled as direct—albeit noisy or incomplete—subsets of the state variables, rather than as functions of an underlying latent regime. Furthermore, while effective for imputation, the core state and observation equations do not explicitly incorporate the network topology into their dynamics. Instead, the topological structure is introduced primarily as an inductive bias during the estimation phase.

To overcome these limitations and effectively handle streaming, incomplete data, there is a critical need for a fully integrated, topology-aware state-space model—one that natively embeds the underlying topological structure into both the state evolution and observation equations, while operating online to dynamically infer latent states.
To address these challenges, we propose a topology-aware state space model defined on cell complexes. Specifically, we formulate the latent state evolution as a linear stochastic partial differential equation (SPDE), where the deterministic drift is governed by Hodge Laplacians to process node, edge, and face signals. Simultaneously, boundary operators propagate uncertainty consistently with the underlying topology. We model observations as nonlinear transformations of these latent states, applying parameterized nonlinear functions before topology-aware convolutions. For efficient online tracking and estimating the parameters, we employ a tailored Expectation Maximization (EM) and Extended Kalman Filter (EKF) framework. Finally, we derive an online heuristic cell identification mechanism to explicitly infer the underlying topological structure from streaming data when higher-order connections are unknown. Our framework generalizes the Graph Kalman Network (GKNet) \cite{sabbaqi2025gknet}. While both approaches share a foundational reliance on stochastic partial differential equations (SPDEs) for state tracking, our work introduces three core methodological advancements: (1) we elevate the underlying modeling domain from standard pairwise graphs to higher-order topological structures; (2) rather than relying on computationally demanding encoder-decoder architectures to handle nonlinearities, we propose an efficient online learning strategy that utilizes Random Fourier Features (RFFs) for lightweight, real-time nonlinear approximation; and (3) we address practical scenarios where the underlying higher-order structures are unknown, introducing a principled mechanism to infer these topological relationships directly from the streaming data.

The main contributions of this paper are:
\begin{itemize}[wide, labelindent=0pt, labelsep=0.5em, nosep]
    \item \textbf{Topological state space modeling with incomplete data:} We propose a state equation that governs latent topological signals via a linear SPDE, using Hodge Laplacians for deterministic drift and the Dirac operator for stochastic diffusion while the observation equation models incomplete data as nonlinear transformations of these states. Furthermore, we provide a spectral analysis demonstrating how this formulation improves observability over purely Laplacian-based methods.
    
    \item \textbf{Joint nonlinear state tracking and parameter learning:} We develop a comprehensive online inference method to solve the nonlinear observation models. To achieve this, we approximate the nonlinear observation mappings using RFFs and derive an online Expectation-Maximization (EM) algorithm that integrates an Extended Kalman Filter (EKF) in the E-step for recursive state tracking, with maximum-likelihood updates in the M-step to dynamically learn the RFF coefficients and system uncertainties in real time.
    
    \item \textbf{Heuristic topological inference:} When only the $1$-skeleton is given along with node and edge signals, we provide a data-driven mechanism to reconstruct missing higher-order topology. Our algorithm iteratively identifies and adds missing $2$-cells by evaluating the NMSE improvements they offer to the state space model, thereby inferring the complete cell complex directly from streaming observations.
\end{itemize}

The paper is organized as follows. Section \ref{S:prelim} introduces the topology-aware state space framework. Section \ref{S:Problem_for} details the online learning algorithm and adaptive cell identification mechanism. Section \ref{S:Frequency} analyzes the system's observability and its behavior in the topological frequency domain. Section \ref{S:exp} presents the experimental evaluation, and Section \ref{S:conclusion} concludes.

\vspace{-0.2cm}
\section{Topology-Aware
State-Space Models} \label{S:prelim}

A state space model consists of: the state equation, which governs the evolution of the latent state, and the observation equation, which relates the latent variable to observations. The state equation describes the evolution of the latent variable $\bbx_t \in \mathbb{R}^N$ over time:
\vspace{-0.2cm}
\begin{equation}\label{eq:state_equation}
\bbx_t = f(\bbx_{t-1}, \boldsymbol{u}_t) + \bbq,
\end{equation}

\noindent where $f(\cdot)$ encodes the system dynamics, $\boldsymbol{u}_t \in \mathbb{R}^m$ represents an input, and $\bbq \sim \mathcal{N}(\mathbf{0}, \bbQ)$ is the process noise with a zero-mean Gaussian distribution with a time-invariant covariance matrix $\bbQ$. The observation equation connects the latent state to the observations $\boldsymbol{y}_t \in \mathbb{R}^{N_o}$:
\vspace{-0.1cm}
\begin{equation}\label{eq:obeserbation_equation}
\boldsymbol{y}_t = g\left(\bbx_t, \bbu_t\right) + \boldsymbol{n},
\end{equation}

\noindent where $g(\cdot)$ maps the latent state into the observation space, and $\boldsymbol{n} \sim \mathcal{N}(\mathbf{0}, \bbR)$ represents Gaussian observation noise.

Applying the general state space formulation in~\eqref{eq:state_equation} and~\eqref{eq:obeserbation_equation} directly to complex physical systems is challenging. Without structural constraints, the transition and observation operators are entirely structure-agnostic. This causes high computational complexity and ignores intrinsic couplings among different orders of signals such as node time-series coupled with edge flows in a critical infrastructure network. To address these issues when the dynamics are defined over topologies~\cite{millan2025topology}, we incorporate the system's underlying topology as an inductive bias. Specifically, we employ cell complexes to model the topology, explicitly integrating higher-order structural dependencies into both the state dynamics and observation models. Before introducing this topology-aware framework, we briefly review cell complexes and their associated signals.

\vspace{-0.2cm}
\subsection{Cell Complex}
\begin{figure}[!t]
    \centering
    \includegraphics[width=0.65\linewidth]{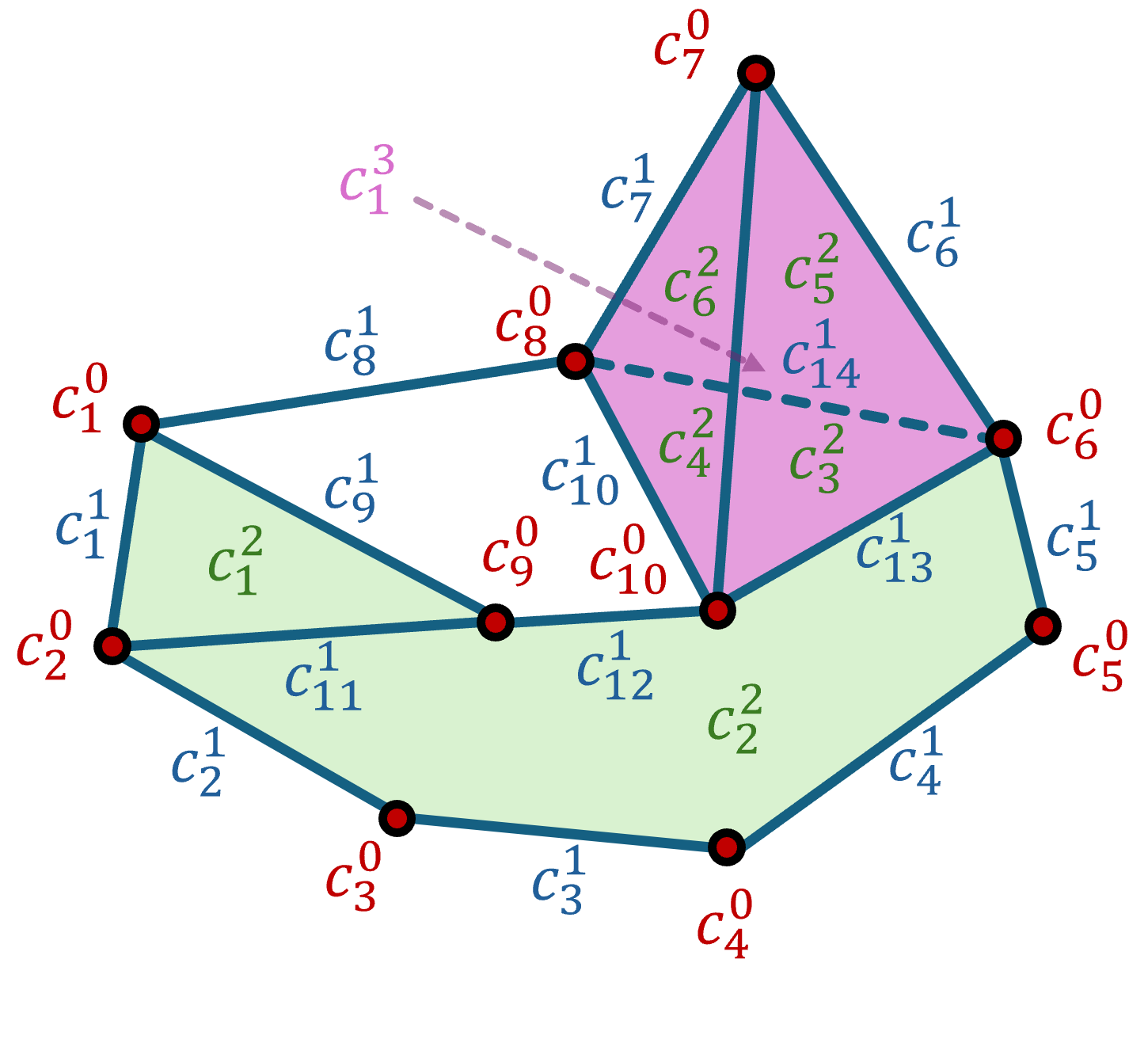}
    \vspace{-0.6cm}
    \caption{A cell complex of order three consisting of ten $0$-cells $c_j^0$, denoted by the red nodes; fourteen $1$-cells $c_j^1$, denoted by the blue lines connecting these nodes; and six $2$-cells $c_j^2$, denoted by the green polytopes. These $2$-cells can be bounded by an arbitrary number of edges, distinguishing them from triangular simplicial faces. Finally, the complex includes one $3$-cell $c_j^3$, denoted by the purple enclosed volumetric region.}
    \label{fig:cell_complex_definition}
\end{figure}
We denote a \textit{cell} by $c_i$, and its order by $\dim(c_i)$. A $k$-\textit{cell} $c_i^k$ is a cell of order $k$ whose underlying topological space is homeomorphic to the open $k-$dimensional Euclidean ball. For instance, a $0-$cell is a single point, a $1-$cell is a line segment, and a $2-$cell is a face. A \textit{regular cell complex} $\mathcal{X}$ is a topological space partitioned into disjoint open cells, where the closure of each $k$-cell is homeomorphic to a closed $k$-ball, and its boundary comprises lower-order cells. A complex is of order $K$, denoted $\mathcal{X}^K$, if its maximum cell order is $K$. For instance, as shown in Fig.~\ref{fig:cell_complex_definition}, a $3$-complex consists of nodes ($0$-cells), edges ($1$-cells), faces ($2$-cells), and polyhedra ($3$-cells).  While a standard graph is merely a $1$-complex limited to pairwise interactions, higher-order complexes (e.g., $K \ge 2$) provide a powerful topological framework to capture complex, multi-way dependencies in both physical and abstract networks.

The structure of a cell complex $\mathcal{X}^{K}$ is fully described by incidence matrices $\mathbf{B}_k$ ($k=1, \ldots, K$), which generalize the standard graph incidence matrix. Let $\mathcal{C}^k$ denote the set of $N_k$ $k$-cells in $\mathcal{X}^{K}$. The matrix $\mathbf{B}_k \in \mathbb{R}^{N_{k-1} \times N_k}$ has entries $[\mathbf{B}_k]_{ij} \in \{-1, 0, 1\}$ that encode the oriented boundary relations between the $k$-cell $c_j^k$ and the $(k-1)$-cell $c_i^{k-1}$. As illustrated in Fig.~1, this sequence captures connectivity across all orders, e.g., nodes to edges for $\mathbf{B}_1$, and edges to faces for $\mathbf{B}_2$. Based on these boundary operators, we define combinatorial \textit{Hodge Laplacian matrices} to capture cellular interactions across orders:

\begin{small}
\begin{equation}\label{eq:hodge_lap}
\left\{\begin{aligned}
& \mathbf{L}_0=\mathbf{B}_1 \mathbf{B}_1^\top, \\
& \mathbf{L}_k=\mathbf{B}_k^\top \mathbf{B}_k+\mathbf{B}_{k+1} \mathbf{B}_{k+1}^\top, k=1, \ldots, K-1, \\
& \mathbf{L}_K=\mathbf{B}_K^\top \mathbf{B}_K.
\end{aligned}\right.
\end{equation}
\end{small}

\noindent Here, $\mathbf{L}_0$ is the graph Laplacian representing node-to-node adjacencies. For the intermediate cells, we have that the $k$-Laplacian $\mathbf{L}_k$ is composed of two terms: $\mathbf{L}_{k,\ell}=\mathbf{B}_k^\top\mathbf{B}_k$ as the lower Laplacian and $\mathbf{L}_{k,u}=\mathbf{B}_{k+1}\mathbf{B}_{k+1}^{\top}$ as the upper Laplacian. An alternative representation of the whole structure is provided by the Dirac operator~\cite{calmon2023dirac} $\mathbf{D}$, which has a block structure where the $(k, k^{\prime})$-th block, denoted by $\llbracket \mathbf{D} \rrbracket_{k, k^{\prime}}$, relates orders $k$ and $k^{\prime}$. Formally, each block is defined as

\begin{small}
\begin{equation} \label{eq:dirac_oper}
\llbracket \mathbf{D} \rrbracket_{k, k^{\prime}} = 
\begin{cases}
\mathbf{B}_{k+1}, & \text{if } k^{\prime} = k+1, \\
\mathbf{B}_{k}^\top, & \text{if } k^{\prime} = k-1, \\
\mathbf{0}, & \text{otherwise,}
\end{cases}
\end{equation}
\end{small}

\noindent with dimension $N=\sum_{k=0}^{K}N_k$. This implies that the Dirac operator $\mathbf{D}$ can also be decomposed into a lower component and an upper component, similar to the Hodge Laplacian, and that $\mathbf{D}^2 = \mathrm{blkdiag}(\mathbf{L}_0, \mathbf{L}_1, \dots, \mathbf{L}_K)$.

\vspace{-0.2cm}
\subsection{Topological Signals}
\textit{Cell signals} $\bbs^k$ are functions mapping from the set of all $k-$cells contained in the complex to real numbers: 
\begin{equation} 
\bbs^k=[s^k(c_1^k), \ldots, s^k(c_i^k), \ldots, s^k(c_{N_k}^k)]\in\mathbb{R}^{N_k}, 
\end{equation}
where $k=0, 1, \ldots, K$. The combination of all cell signals of different orders forms the complete cell complex signal $\bbs=[(\bbs^{0})^\top, (\bbs^{1})^\top, \dots, (\bbs^{K})^\top]^\top\in\mathbb{R}^{N}$. For example, in Fig.~\ref{fig:cell_complex_definition} with $\mathcal{X}^{3}$, $\bbs^0\in\mathbb{R}^{N_0}$, $\bbs^1\in\mathbb{R}^{N_1}$, $\bbs^2\in\mathbb{R}^{N_2}$, and $\bbs^3\in\mathbb{R}^{N_3}$ are node, edge, face, and polyhedral signals, respectively. These signals admit tangible physical interpretations. In water networks, $\bbs^0$ corresponds to water pressure at junctions, while $\bbs^1$ quantifies the flow through pipes. Crucially, the inclusion of $2-$cells (faces) explicitly models topological loops, capturing the conservation laws~\cite{schaub2021signal}. Analogously, in traffic networks, faces represent city blocks, encoding regional congestion patterns or circular dynamics essential for macroscopic analysis.

Formally, the signal-topological coupling is mediated by the incidence matrices ($\mathbf{B}_k$) and their adjoints ($\mathbf{B}_k^\top$). Focusing on the edge signal $\bbs^{1} \in \mathbb{R}^{N_1}$, these matrices act as operators that transform signals vertically between cell orders:

\begin{itemize}[wide, labelindent=0pt, labelsep=0.5em, nosep]

    \item \textit{Gradient operator} $\mathbf{B}_1^\top$: transforms node signals to edge signals ($\mathbb{R}^{N_0} \to \mathbb{R}^{N_1}$) via $\operatorname{grad}(\bbs^{0}) = \mathbf{B}_1^\top \bbs^{0}$. Mathematically, $\mathbf{B}_1^\top$ computes the difference between the signal values of the two endpoints for each edge. Physically, this captures edge flows induced by node potential gradients (e.g., flow driven by pressure differences).

    \item \textit{Divergence operator} $\mathbf{B}_1$: transforms edge signals to node signals ($\mathbb{R}^{N_1} \to \mathbb{R}^{N_0}$) via $\operatorname{div}(\bbs^{1})=\mathbf{B}_1 \bbs^{1}$. The matrix sums the signed flows of all edges incident to a specific node. The resulting node signal measures the flow divergence at nodes.

    \item \textit{Curl operator} $\mathbf{B}_2^\top$: transforms edge signals to face signals ($\mathbb{R}^{N_1} \to \mathbb{R}^{N_2}$) via $\operatorname{curl}(\bbs^{1})=\mathbf{B}_2^{\top} \bbs^{1}$. It sums the flows of all edges along the boundary of a specific face. The resulting face signal quantifies the total circulation around the face.

    \item \textit{Curl Adjoint operator} $\mathbf{B}_2$: transforms face signals to edge signals via $\operatorname{curl}^*(\bbs^{2})=\mathbf{B}_2 \bbs^{2}$. It projects a signal defined on a face onto its boundary edges, thereby representing the rotational flow contribution induced by the face.

\end{itemize}

Distinct from the inter-order mappings of incidence matrices, the Hodge Laplacians facilitate intra-order signal propagation. When applied to an edge signal $\bbs^1$, the lower Laplacian $\mathbf{L}_{1,\ell}=\mathbf{B}_1^\top \mathbf{B}_1$ acts as a shift operator via the mapping $\bbs^1 \mapsto \mathbf{L}_{1,\ell}\bbs^1$, which physically diffuses the signal among edges that share a common node. Similarly, the upper Laplacian $\mathbf{L}_{1,u}=\mathbf{B}_2 \mathbf{B}_2^\top$ governs diffusion via upper adjacency through the mapping $\bbs^1 \mapsto \mathbf{L}_{1,u}\bbs^1$, shifting the signal between edges that bound the same face. The total Hodge Laplacian $\mathbf{L}_1$ integrates these two operations, enabling simultaneous signal mixing across adjacent cells sharing either a common lower-order node or a higher-order face.

{\vspace{-0.4cm}
\subsection{State Space Model for Cell Complex}
We now propose a topology-aware state space formulation for cell complex signals.}

\subsubsection{State Model} We employ a structure-aware state formulation grounded in SPDEs to model the underlying system dynamics. Such formulations offer a versatile description of time-varying processes encountered in diverse applications~\cite{sarkka2019applied}. A canonical example is the stochastic heat equation, which is:

\vspace{-0.2cm}
{\small{\begin{equation}
\frac{d \bbx_t}{dt} = c \nabla^2 \bbx_t + \bbzeta_t.
\end{equation}}}

\noindent The dynamics of process $\bbx_t$ is determined by the Laplacian operator $\nabla^2$, where the stochastic white noise $\bbzeta_t \in \mathbb{R}^N$ introduces randomness, and the diffusion coefficient $c$ controls the rate of propagation. To fuse the cell structure information into the evolution of the process, the continuous Laplace operator $\nabla^2$ is replaced by its discrete analogue~\cite{wardetzky2007discrete,sabbaqi2025gknet}, yielding the unified Stochastic Differential Equation (SDE):

\vspace{-0.2cm}
{\small{\begin{equation}\label{eq:dicre_spde}
d \bbx_t=-c \bbL \bbx_t d t+\bbS d \bbbeta_t
\end{equation}}}

\noindent Crucially, $\bbL$ denotes the joint Hodge Laplacian defined as the block-diagonal matrix $\bbL = \operatorname{blkdiag}(\bbL_0, \bbL_1, \dots, \bbL_K)$, which unifies the diffusion dynamics across all cell orders. Here, $\bbbeta_t \in \mathbb{R}^N$ represents a standard Brownian motion, which serves as a driving force mediated by the dispersion matrix $\bbS \in \mathbb{R}^{N \times N}$. The matrix $\bbS$ determines how uncertainty is introduced into the system. According to Equation~\eqref{eq:dicre_spde}, the evolution at each cell order is effectively controlled by its corresponding block $\bbL_k$ within the global operator $\bbL$, subject to inherent uncertainty that is coupled across all orders as $\bbS$ is still topology-agnostic. Signal $\bbx_t$ is the concatenation of signals across all cell orders.

\begin{remark}
Under Gaussian initial conditions with mean $\boldsymbol{\mu}_0$, the solution to \eqref{eq:dicre_spde} is a Gaussian Process $\bbx_t \sim \mathcal{GP}(\boldsymbol{\mu}_t, \mathbf{K}_{t,s})$, where the mean function evolves as $\boldsymbol{\mu}_t = \exp(-c \bbL t)\boldsymbol{\mu}_0$. Directly utilizing this GP formulation presents two fundamental limitations. First, evaluating the spatiotemporal covariance kernel $\mathbf{K}_{t,s}$ involves an eigen-decomposition of $\bbL$, leading to a prohibitive computational burden scaling as $\mathcal{O}(N^3T^3)$, where $T$ denotes the number of time steps~\cite{sabbaqi2025gknet}. Second, a generic dispersion matrix $\bbS$ fails to align the stochastic forcing with the underlying topological connectivity. To overcome these issues, we transition to a discrete state space formulation incorporating a topology-aware dispersion matrix.
\end{remark}

We constrain the dispersion matrix $\bbS$ by modeling how uncertainty propagates through the cell complex. Physically, uncertainty transmits via boundary operators between adjacent cell orders, akin to fluxes in conservation laws. Thus, perturbations in node signals $\bbx_t^0$ (governed by $\bbL_0$) originate from their incident edges. Similarly, edge signals are perturbed by adjacent nodes and bounding faces, while face signals are affected by their boundary edges. Based on this incidence-driven rationale, we reformulate~\eqref{eq:dicre_spde} as follows:

{\small{
\begin{equation}\label{eq:dirac_spade}
d \bbx_t=-c \bbL \bbx_t dt+\bbD \operatorname{diag}(\bbalpha) d \bbbeta_t
\end{equation}}}

The Dirac operator $\bbD$ unifies cross-order interactions. Its off-diagonal blocks contain boundary operators $\bbB_k$ and their adjoints $\bbB_k^\top$, which naturally propagate stochastic perturbations across adjacent orders. For example, edge-level uncertainty projects onto connected nodes via $\bbB_1$ and onto faces via $\bbB_2^\top$. Finally, the vector $\boldsymbol{\alpha} \in \mathbb{R}^N$ in~\eqref{eq:dirac_spade} quantifies the uncertainty for each cell signal. Instead of being a fixed prior, $\boldsymbol{\alpha}$ will be inferred jointly with other model parameters.

\begin{remark}[Discrete state equation]
Assuming high-resolution uniform sampling with interval $\delta t$, the continuous state equation is discretized via a first-order approximation~\cite{sabbaqi2025gknet}:

{\small{\begin{equation}
    \bbx_{t_i+\delta t} = \tilde{\bbL}\, \bbx_{t_i} + \bbq_i, 
    \quad \bbq_i \sim \mathcal{N}(0, \bbQ),
\end{equation}}}
where the transition operator $\tilde{\bbL}$ and noise covariance $\bbQ$ are given by~\cite{sabbaqi2025gknet}:
{\small{
\begin{equation}\label{eq:discre_state}
\begin{cases}
\begin{aligned}
    \tilde{\bbL} &\simeq \bbI - c \delta t\, \bbL,\\
    \bbQ &\simeq \delta t\, \bbD\, \mathrm{diag}^2(\boldsymbol{\alpha})\, \bbD^\top.
\end{aligned}
\end{cases}
\end{equation}}}
Based on~\eqref{eq:discre_state}, $\bbQ$ depends on both the uncertainty vector $\boldsymbol{\alpha}$ and the topological Dirac operator $\bbD$. This topology-aware formulation reduces the required noise parameters, achieving linear scalability for uncertainty parameters with respect to the latent states~\cite{money2024kalman}.
\end{remark}

While $\bbQ$ is theoretically singular, in practice, numerical stability can be achieved by regularizing the covariance matrix as $\bbQ + \gamma\mathbf{I}$ ($\gamma > 0$) in~\eqref{eq:discre_state}. 

\subsubsection{Observation Model}
Building on the structure-aware dynamics in~\eqref{eq:dicre_spde}, let $t_i$ denote the specific time instances at which the system is evaluated. For notational simplicity, we use the discrete index $i \in \{1, 2, \dots\}$ to represent time step $t_i$, and define the discrete latent state as $\bbx_i \equiv \bbx(t_i)$. At each step $i$, we obtain a noisy, nonlinear observation of $\bbx_i$ as follows:
\vspace{-0.1cm}
\begin{equation}\label{eq:observation}
\boldsymbol{y}_i = \boldsymbol{\Phi}\mathcal{M}\left(\bbx_i + f\left(\bbx_i\right)\right) + \boldsymbol{n}_i,
\end{equation}
where the sampling matrix $\bbPhi\in \{0,1\}^{N_o \times N}$ selects $N_o$ observed entries, $\bbPhi = \mathbf{I}$ denotes complete observation, and $\boldsymbol{n}_i \in \mathbb{R}^{N_o}$ is the observation noise. The generic linear operator $\mathcal{M}$, models multi-hop topological signal aggregation of the state. Finally, to increase expressivity beyond purely linear interactions,  function $f(\bbx_i)$ captures complex nonlinear relationships between the latent states and the observations.

We explicitly separate the linear and nonlinear components to adopt a \textit{residual learning} strategy, even though a generic nonlinear mapping could theoretically capture all dynamics. The linear term models dominant baseline trends, while the nonlinear function $f(\cdot)$ solely captures complex deviations. This division reduces the approximation burden on $f(\cdot)$, thereby improving optimization efficiency and numerical stability over single black-box models. This decomposition aligns with Residual Nonlinear Estimators (ResNEsts)~\cite{chen2021resnests}, which has shown potential to outperform the optimal linear predictor.

To model interactions across arbitrary topological orders in a unified manner, we instantiate the linear mapping operator $\mathcal{M}$ as a topological filtering operator $\bbH(\bbL) \in \mathbb{R}^{N \times N}$ that acts jointly on the concatenated signals of a general cell complex of order $K$. This operator captures both intra-order diffusion (within each topological order) and inter-order coupling (between adjacent orders) through the boundary operators $\bbB_k$. Similar to the block structure of the Dirac operator in Eq.~\eqref{eq:dirac_oper}, the operator $\mathcal{M}$ can be defined by its $(k, k^{\prime})$-th block as:

\begin{small}
\begin{equation} \label{eq:H_operator}
\llbracket \mathcal{M} \rrbracket_{k, k^{\prime}} = 
\begin{cases}
\bbH(\bbL_k), & \text{if } k^{\prime} = k, \\
\bbB_{k^{\prime}} \bbH(\bbL_{k^{\prime}}), & \text{if } k^{\prime} = k+1, \\
\bbB_k^\top \bbH(\bbL_{k^{\prime}}), & \text{if } k^{\prime} = k-1, \\
\bb0, & \text{otherwise.}
\end{cases}
\end{equation}
\end{small}

\noindent where each block filter $\bbH(\bbL_k)$ governs the information within the $k-$th topological order. Each intra-order filter $\bbH(\bbL_k)$ is parameterized as a polynomial function of the corresponding Laplacian, following the topological convolution ~\cite{yang2022simplicial}:

\begin{small}
\begin{equation}
\bbH(\bbL_k)
= \sum_{n=0}^{L_{\ell}-1} h_{n,\ell} \, \bbL_{k,\ell}^n
+ \sum_{n=0}^{L_u-1} h_{n,u} \, \bbL_{k,u}^n,
\label{eq:H_polynomial}
\end{equation}
\end{small}

\noindent where $\bbL_{k,\ell}$ and $\bbL_{k,u}$ denote the lower and upper components of the $k-$th Hodge Laplacian, respectively~\cite{yang2022simplicial}. 
The associated coefficients $\mathbf{h}_\ell = \{h_{n,\ell}\}_{n=0}^{L_\ell-1} \in \mathbb{R}^{L_\ell}$ and 
$\mathbf{h}_u = \{h_{n,u}\}_{n=0}^{L_u-1} \in \mathbb{R}^{L_u}$ are the filter taps governing multi-hop cell complex signal propagation. For simplicity, the overall parameter set is given by $\mathbf{h} = \{\mathbf{h}_\ell, \mathbf{h}_u\}$.

This decoupled parameterization is justified by the fundamental topological property $\mathbf{B}_{k}\mathbf{B}_{k+1} = \mathbf{0}$, which implies the orthogonality of the constituent Laplacians (i.e., $\mathbf{L}_{k,\ell}\mathbf{L}_{k,u} = \mathbf{0}$). Due to this property, the cross-terms in the polynomial expansion vanish, satisfying $\mathbf{L}_k^n = \mathbf{L}_{k,\ell}^n + \mathbf{L}_{k,u}^n$. Physically, this independence ensures that the filter $\bbH(\bbL_k)$ performs two distinct \textit{multi-hop} aggregation processes without interference: the first term in~\eqref{eq:H_polynomial} diffuses information exclusively by reaching $n$-hop neighbors via lower adjacencies (e.g., edges communicating through shared nodes), while the second propagates signals via upper adjacencies (e.g., edges communicating through shared faces). By exploiting this separation, our formulation allows for learning specific interaction ranges ($L_\ell$ versus $L_u$) for the lower and upper topological pathways.

For brevity, we denote $\bbH(\bbL)$ by $\bbH$ and its parameter vector by $\mathbf{h}$. The following section introduces an optimization strategy for jointly estimating the evolving states and the unknown parameters in ~\eqref{eq:observation}.

\section{INFERENCE AND LEARNING}\label{S:Problem_for}
This section presents the core algorithms of our framework. We first formulate the learning setup for joint state tracking and parameter estimation under partial observability. We then employ an extended Kalman filter for online state inference, coupled with an expectation-maximization algorithm to learn the unknown model parameters. Finally, we propose a data-driven heuristic to explicitly identify the underlying higher-order structures, specifically the active $2$-cells.
\subsection{Learning Setup and Objectives}
We consider dynamical systems on cell complexes. Although our formulation generalizes to any order $K$, we focus on $K=2$ to ease exposition and because of its versatility in modeling practical multi-way interactions. In this setting, the base $1$-skeleton (nodes and edges) is fully known, while the active $2$-cells driving the latent dynamics may be unknown. Let $\{\boldsymbol{y}_i\}_{i=1}^{T}$, with $\boldsymbol{y}_i\in\mathbb{R}^{N_o}$, denote a stream of noisy, partial observations arriving sequentially, where $T$ denotes the current time index and increases as the stream continues.

We model the latent topological state $\bbx_i \in \mathbb{R}^N$ and the observations via the following discrete-time stochastic state-space model:

\vspace{-0.2cm}
{\small{
\begin{equation} \label{eq:state_space}
\begin{cases}
\bbx_i = \tilde{\mathbf{L}}\bbx_{i-1} + \mathbf{q}_i, \\[3pt]
\boldsymbol{y}_i = \boldsymbol{\Phi}\mathbf{H}(\mathbf{L})\big(\bbx_i + f(\bbx_i)\big) + \mathbf{n}_i,
\end{cases}
\end{equation}
}}

\noindent where $\mathbf{q}_i \sim \mathcal{N}(\mathbf{0}, \mathbf{Q}(\boldsymbol{\alpha}))$ and $\mathbf{n}_i \sim \mathcal{N}(\mathbf{0}, \mathbf{R})$ represent the zero-mean Gaussian process and measurement noise terms, respectively. We model as $\mathbf{R}$ as a diagonal matrix $\mathbf{R} = \sigma^2\mathbf{I}$. Crucially, the process noise covariance is parameterized via the Dirac operator $\mathbf{D}$ as:

\vspace{-0.2cm}
\begin{equation} \label{eq:Q_alpha}
\mathbf{Q}(\boldsymbol{\alpha}) = \delta_t \mathbf{D} \mathrm{diag}^2(\boldsymbol{\alpha}) \mathbf{D}^\top,
\end{equation}

where $\boldsymbol{\alpha}$ is a vector that explicitly models the uncertainty across the cell complex. The operator $\mathbf{H}(\mathbf{L})$ serves as a topology-aware filter~\cite{yang2022simplicial}, $f(\cdot)$ captures nonlinear effects, and $\boldsymbol{\Phi} \in \{0, 1\}^{N_o \times N}$ is a sampling matrix selecting the observed components.

Our online  problem targets three coupled goals:
\begin{itemize}[wide, labelindent=0pt, labelsep=0.5em, nosep]
\item \textbf{Latent state estimation:} track $\bbx_i$ for forecasting, denoising, and imputation under partial observability.
\item \textbf{Parameter learning:} learn uncertainty parameters $\bbalpha$, convolution filter coefficients defining $\bbH(\bbL)$, and the nonlinear mapping $f(\cdot)$.
\item \textbf{Topology inference:} when only the graph structure is known, infer which candidate faces are active and necessary to explain the stream.
\end{itemize}

We address online inference under two structural settings. First, we perform joint state and parameter learning from streaming, partial observations given the full cell complex structure as an inductive bias. To enable efficient online updates, we approximate the Reproducing Kernel Hilbert Space (RKHS) nonlinearity $f(\cdot)$ using RFFs. We then apply an Expectation-Maximization (EM) framework: the E-step recursively tracks latent states via an Extended Kalman Filter (EKF), while the M-step updates model parameters online using maximum likelihood.

Second, we address scenarios where given a structure of order $K$ and respective time-series, we aim to also infer the cells of order $K+1$ in an online manner. More specifically, given a $1$-skeleton, we augment the EKF-EM recursion with an online cell identification module. This is particularly relevant in critical infrastructure networks where the 1-skeleton is given and the 2-cells are unknown and are to be retrieved from the node and edge time-series as a way to identify hidden structures governing the dynamics. This module sequentially activates candidate $2$-cells that improve predictive performance, yielding a parsimonious topology. We first detail the RFF parameterization for $f(\cdot)$.
{\vspace{-.3cm}\subsection{Nonlinear Function Parameterization}
The observation model in~\eqref{eq:state_space} considers a nonlinearity $f(\bbx_i)$ to capture localized nonlinear sensor physics. To retain expressive modeling while enabling online learning, we parameterize $f(\cdot)$ in a Reproducing Kernel Hilbert Space (RKHS) and then adopt a fixed-dimensional random-feature (RFF) approximation to replace the growing kernel expansion with a fixed-dimensional parametrization. The resulting measurement model remains nonlinear in the latent state, and we handle this nonlinearity via EKF linearization in the next.
}

\subsubsection{Kernel-based point-wise nonlinearity}
We assume $f(\bbx_i)$ acts entrywise, i.e.,
$f(\bbx_i)=[f(x_i[1]),\ldots,f(x_i[N])]^\top$, capturing localized nonlinear responses.
Global coupling across cells is captured by the topology-aware filter $\bbH(\bbL)$. Let $f(\cdot)\in\mathcal{H}_\kappa$ be an RKHS induced by a positive-definite kernel $\kappa(\cdot,\cdot)$.
By the representer theorem~\cite{scholkopf2001generalized}, the $n$th component admits the expansion:

\vspace{-0.3cm}
{\small{
\begin{equation}\label{eq:rkhs_expansion}
\hat f(x_i[n])=\sum_{i'=1}^{N_s}W_{i',n}\,\kappa\!\big(x_i[n],x_{i'}[n]\big),
\end{equation}}}

where $\{x_{i'}[n]\}_{i'=1}^{N_s}$ are past samples and $W_{i',n}$ are coefficients.
While expressive,~\eqref{eq:rkhs_expansion} is unsuitable for streaming since the number of parameters grows with the number of observed samples $N_s$.

\subsubsection{Random Fourier feature approximation}
To obtain a fixed-dimensional representation, we approximate the shift-invariant kernel using RFFs. Let $x_i[n]$ and $x_{i'}[n]$ be the inputs. By Bochner's theorem~\cite{bochner1959lectures}, any shift-invariant kernel $\kappa$ is the Fourier transform of a probability density function $\pi_\kappa$. By defining $v \sim \pi_\kappa$ as the sampled spectral frequency, the kernel becomes:
\vspace{-0.1cm}
\begin{equation}\label{eq:rf_approximation}
\kappa(x_i[n],x_{i'}[n])
=\mathbb{E}_{v\sim\pi_\kappa}\!\left[e^{jv(x_i[n]-x_{i'}[n])}\right].
\end{equation}
Drawing $\{v_m\}_{m=1}^M\sim\pi_\kappa$ yields the feature map
\vspace{-0.2cm}
\begin{equation}\label{eq:feature_vector}
\begin{aligned}
\boldsymbol z_v(x_i[n]) \triangleq \frac{1}{\sqrt{M}} \big[ & \sin(v_1x_i[n]),\ldots,\sin(v_Mx_i[n]), \\
& \cos(v_1x_i[n]),\ldots,\cos(v_Mx_i[n]) \big]^\top,
\end{aligned}
\end{equation}
so that $\kappa(x_i[n],x_{i'}[n])\approx \boldsymbol z_v(x_i[n])^\top \boldsymbol z_v(x_{i'}[n])$.
Substituting this approximation into~\eqref{eq:rkhs_expansion} gives
{\small{
\begin{equation}\label{eq:approximation_nonlinear_rewrite}
\hat f(x_i[n]) \approx \bbGamma_n\,\boldsymbol z_v(x_i[n]),\qquad \bbGamma_n\in\mathbb{R}^{1\times 2M}.
\end{equation}}}

\noindent Hence, $\hat f(x_i[n])$ is linear in the parameters $\bbGamma_n$ and with a fixed dimension $2M$, but remains nonlinear in the state through $\boldsymbol z_v(x_i[n])$.
In the following, we collect $\{\bbGamma_n\}_{n=1}^N$ into $\bbGamma\in\mathbb{R}^{N\times 2M}$ and treat $\bbGamma$ as learnable parameters that will be learned online jointly with the filter parameters.
{\vspace{-.3cm}\subsection{Online Approximated Expectation--Maximization}
Having defined the nonlinear measurement function:}
{\small{
\begin{equation}
\boldsymbol{y}_i = \bbPhi\,\bbH(\bbL)\big(\bbx_i + f(\bbx_i)\big) + \bbn_i
\end{equation}}}

\noindent we perform joint state inference and parameter learning via an online EM-type recursion.
At time $i$, the E-step approximates the posterior of $\bbx_i$ given current parameters, and the M-step updates $(\bbalpha,\bbh,\bbGamma)$ by maximizing the resulting approximate conditional likelihood.

\subsubsection{E-step}
 We employ an Extended Kalman Filter (EKF) to track the state evolution based on the underlying dynamical model and obtained observations.
 
\noindent\textbf{Prediction:}
{\small
\begin{empheq}[left = \empheqlbrace]{align}
\bbx_{i|i-1} &= \tilde{\bbL}\bbx_{i-1|i-1}, \label{eq:pre1} \\
\bbP_{i|i-1} &= \tilde{\bbL}\bbP_{i-1|i-1}\tilde{\bbL}^\top + \bbQ(\bbalpha). \label{eq:pre2}
\end{empheq}}

\noindent Here, $\bbx_{i|i-1}$ and $\bbP_{i|i-1}$ are the a priori state estimate and error covariance at step $i$, $\bbx_{i-1|i-1}$ and $\bbP_{i-1|i-1}$ are the a posteriori updates from step $i-1$, and $\bbQ(\bbalpha)$ is the parameterized process noise covariance.

\noindent\textbf{Correction:}
Linearize $\hat{\boldsymbol{y}}(\bbx)$ at the prior state estimate $\bbx_{i|i-1}$:

{\small{\begin{equation}
\bbJ_i \triangleq \frac{\partial \hat{\boldsymbol{y}}(\bbx)}{\partial \bbx}\Big|_{\bbx_{i|i-1}}
=\bbPhi\,\bbH(\bbL)\big(\bbI+\bbG(\bbx_{i|i-1})\big),
\end{equation}}}

where $\bbG(\bbx) \triangleq \mathrm{diag}(\partial \hat f(x[1]) / \partial x[1], \dots, \partial \hat f(x[N]) / \partial x[N])$. With~\eqref{eq:approximation_nonlinear_rewrite}, each diagonal entry is available in closed form:
\begin{equation}
\partial \hat f(x[n]) / \partial x[n] = \bbGamma_n\,\partial \boldsymbol z_v(x[n]) / \partial x[n],
\end{equation}
and $\partial \boldsymbol z_v(x[n])/\partial x[n]$ follows directly from~\eqref{eq:feature_vector}.
Then

{\small
\begin{empheq}[left=\empheqlbrace]{align}
&\mathbf{S}_i = \mathbf{J}_i\mathbf{P}_{i|i-1}\mathbf{J}_i^\top + \mathbf{R} \label{eq:corr1}\\
&\mathbf{K}_i = \mathbf{P}_{i|i-1}\mathbf{J}_i^\top \mathbf{S}_i^{-1} \label{eq:corr2} \\
&\bbx_{i|i} = \bbx_{i|i-1} + \mathbf{K}_i\big(\boldsymbol{y}_i-\hat{\boldsymbol{y}}(\bbx_{i|i-1})\big) \label{eq:corr3} \\
&\mathbf{P}_{i|i} = (\mathbf{I}-\mathbf{K}_i\mathbf{J}_i)\mathbf{P}_{i|i-1} \label{eq:corr4}
\end{empheq}}

\noindent where $\mathbf{K}_i$ is the Kalman gain, and $\mathbf{R}$ denotes the observation noise covariance matrix.

\subsubsection{M-step: online parameter updates}
Given the EKF quantities, we model the predictive likelihood of the current observation as a Gaussian distribution:
\vspace{-0.2cm}
\begin{equation}
p(\boldsymbol{y}_i\mid \boldsymbol{y}_{1:i-1}) \approx \mathcal{N}\big(\boldsymbol\mu_i,\bbS_i\big),
\qquad \boldsymbol\mu_i \triangleq \hat{\boldsymbol{y}}(\bbx_{i|i-1}).
\end{equation}
The instantaneous negative log-likelihood is
{\small{
\begin{equation}\label{eq:loss_function_rewrite}
\ell_i(\bbalpha,\bbh,\bbGamma)
=\tfrac{1}{2}\log|\bbS_i|
+\tfrac{1}{2}\big(\boldsymbol{y}_i-\boldsymbol\mu_i\big)^\top \bbS_i^{-1}\big(\boldsymbol{y}_i-\boldsymbol\mu_i\big).
\end{equation}}}
We then perform online updates for the model parameters $(\bbalpha, \bbh, \bbGamma)$, specifically, the process noise parameters $\bbalpha$, the convolution filter coefficients $\bbh$, and the nonlinear RFF weights $\bbGamma$, via stochastic gradient steps or alternative optimization schemes:
\vspace{-0.2cm}

{\small{\begin{empheq}[left=\empheqlbrace]{align}
\bbalpha_i &= \bbalpha_{i-1} - a_1 \nabla_{\bbalpha}\ell_i, \label{grad1} \\
\bbh_i      &= \bbh_{i-1}      - a_2 \nabla_{\bbh}\ell_i,     \label{grad2} \\
\bbGamma_i &= \bbGamma_{i-1} - a_3 \nabla_{\bbGamma}\ell_i, \label{grad3}
\end{empheq}}}

where $a_1,a_2,a_3$ are learning rates.
These updates allow the model to continuously refine uncertainty, filtering, and nonlinear parameters as new data arrive.

\subsubsection{Computational complexity}
The computational complexity is dominated by the E-step. The per-time-step complexity is dominated by the propagation and update of the state covariance matrix $\mathbf{P}\in\mathbb{R}^{N\times N}$ and the Kalman-gain computation. Concretely, the covariance propagation $\tilde{\mathbf{L}}\mathbf{P}\tilde{\mathbf{L}}^\top$ and the subsequent covariance updates entail dense matrix multiplications costing $\mathcal{O}(N^3)$ and inverting the innovation covariance contributes up to $\mathcal{O}(N^3)$. Hence, for a sequence of length $T$ the overall worst-case computational complexity of the E-step is $\mathcal{O}(T N^3)$, and the corresponding memory requirement is $\mathcal{O}(N^2)$ due to storage of the full covariance $\mathbf{P}$. 
\begin{algorithm}
\caption{TKF: Topological Kalman Filter}
\label{alg:TKF}
\footnotesize{
\begin{algorithmic}[1]
    \STATE \textbf{Output:} $\bbx_{i|i}$, $\bbh_i$, $\bbalpha_i$, $\bbGamma_i$
    \STATE \textbf{Given:} $\bbB_1$, $\bbB_2$
    \STATE \textbf{Initialize:} $a_1$, $a_2$, $a_3$, $c$, $\delta t$, $\bbx_{0|0}$, $\bbh_0$, $\bbalpha_0$, $\bbGamma_0$
    \FOR{$i = 1$ \TO $T$}
        \STATE Obtain the observation $\boldsymbol{y}_i$
        \STATE Prediction steps~\eqref{eq:pre1}~\eqref{eq:pre2} to obtain $\bbx_{i|i-1}$
        \STATE Correction steps~\eqref{eq:corr1}~\eqref{eq:corr2}~\eqref{eq:corr3}~\eqref{eq:corr4} to obtain $\bbx_{i|i}$
        \STATE Update $\bbalpha_i$, $\bbh_i$,  $\bbGamma_i$ by~\eqref{grad1},~\eqref{grad2},~\eqref{grad3}
    \ENDFOR
\end{algorithmic}
}
\end{algorithm}

\vspace{-0.5cm}
{\subsection{Online cell identification with State Rollback}
We propose a modular $2$-cell identification mechanism, deployed only when inferring relevant higher-order interactions is deemed necessary for downstream tasks. We propose a two-step procedure: given the $1$-skeleton, streaming observations of length $T$, and a pool of $N_2$ candidate $2$-cells as inputs. Crucially, this candidate pool is generated by extracting simple cycles from the $1$-skeleton. The algorithm first considers a warm-up phase of $T_s$ steps to allow the filter's state estimates to stabilize, and then proceeds to the cell identification phase. For the latter, we partition the remaining time-series evenly into $N_2$ discrete windows of $(T-T_s)/N_2$ steps.

In each window, the algorithm temporarily activates one candidate prioritized by estimated uncertainty and executes EKF updates using this candidate $2$-cell over a test window. Crucially, evaluating an incorrect topological structure over a time window inevitably corrupts the state estimate $\bbx$ and error covariance $\bbP$, as well as the learned parameters. To prevent this state leakage from propagating to subsequent evaluations, we implement a strict snapshot-and-rollback mechanism. If the activated cell sufficiently reduces the forecasting NMSE, it is permanently retained. Otherwise, the model discards the $2$-cell, rolls back all EKF variables to their pre-window states, and re-runs the EKF in the window using the input topology without the candidate cell.

\subsubsection{Uncertainty-driven scoring of candidate cells}
To pick candidate $2$-cells, we leverage the learned process uncertainty as a proxy for missing higher-order constraints. Recall the topology-aware approximation of the process covariance:
\begin{equation}
    \bbQ \simeq \delta_t\,\bbD\,\mathrm{diag}^{2}(\bbalpha)\,\bbD^{\top}, \nonumber
\end{equation}
where $\bbalpha = [\bbalpha^{(0)\top}, \bbalpha^{(1)\top}, \bbalpha^{(2)\top}]^{\top}$ encodes localized uncertainties across cells of different orders. That is, entry $\bbalpha^{(k)}$ captures the uncertainty and dynamic noise variance specific to the $k$-order cells.

Intuitively, if the topology lacks a required higher-order cell, the EKF attributes large uncertainty to the involved edges. Without the structure to model these multi-way correlations, the filter is forced to inflate localized edge uncertainties to absorb the resulting forecasting NMSEs. For a candidate $2$-cell $p$ with boundary edge set $\mathcal{E}_{p}$, we define the score using the edge uncertainties $\bbalpha^{(1)}$:

{\small{\begin{equation}\label{eq:uncer}
    \xi_{k}=\frac{1}{|\mathcal{E}_{k}|}\sum_{j\in\mathcal{E}_{k}}\big(\alpha_{j}^{(1)}\big)^{2},
\end{equation}}}

\noindent and collect $\bbxi=[\xi_{1},\ldots,\xi_{N_{2}}]^{\top}$. A larger score $\xi_k$ indicates a loop whose boundary edges remain uncertain under the current model, suggesting that activating cell $p$ (via $\bbB_{2}$) may introduce the missing higher-order coupling and reduce forecasting NMSE.

\subsubsection{Performance-driven cell identification}
Let $\bbe\in\{0,1\}^{N_{2}}$ denote the $2$-cell activation vector and define
\begin{equation}
    \bbB_{2}=\bbB_{2}^{*}\,\mathrm{diag}(\bbe),
\end{equation}
where $\bbB_{2}^{*}$ enumerates all candidate $2$-cells consistent with the known edge set. Crucially, the dimensions of $\bbB_{2}$, along with the associated Dirac operator, state vector, and covariance matrix, remain fixed to the size of the full candidate pool. The inclusion of candidate $2$-cells is controlled by a binary mask vector $\bbe$. An inactive $2$-cell (entry $0$) zeros out its corresponding column in the full incidence matrix $\bbB_{2}^*$, removing its topological influence without resizing the EKF variables. 

Since continuously optimizing $\bbe$ introduces a computationally prohibitive combinatorial loop alongside the estimation of $(\bbx_i,\bbalpha,\bbh,\bbGamma)$, we adopt a forward-selection alternative augmented with a state-preservation protocol. During the identification phase, candidates are sequentially tested based on $\bbxi$ and permanently accepted only if they reduce the NMSE.

\begin{algorithm}[t]
\caption{Online Cell Identification with State Rollback}
\label{alg:online_cell}
{\footnotesize
\begin{algorithmic}[1]
    \STATE \textbf{Input:} Total iterations $T$, warm-up length $T_s$, full incidence matrix $\bbB_2^*$, number of 2-cells $N_2$
    \STATE \textbf{Output:} Updated incidence matrix $\bbB_2$
    \STATE Initialize $\bbe = \boldsymbol{0}$ and set $\bbB_2 = \bbB_2^* \mathrm{diag}(\bbe)$
    \STATE Run Algorithm 1 for $T_s$ steps without any $2$-cells
    \STATE Compute initial forecasting $\text{NMSE}_0$ and set $\text{NMSE}_{\min} = \text{NMSE}_0$
    \STATE Estimate uncertainties $\bbxi$ based on $\bbalpha$ following~\eqref{eq:uncer}
    \FOR{$p = 1$ \TO $N_2$}
        \STATE Identify the $2$-cell with the $p$-th largest $\xi_{i^{(p)}}$
        \STATE Cache current states $(\bbx, \bbP)$ and $(\bbalpha, \bbh, \bbGamma)$
        \STATE $\bbe(p) \gets 1$, update $\bbB_2 \gets \bbB_2^* \mathrm{diag}(\bbe)$
        \FOR{$i = T_s + (p{-}1) \frac{T{-}T_s}{N_2}$ \TO $T_s + p \frac{T{-}T_s}{N_2}$}
            \STATE Update parameters via~\eqref{grad1}--\eqref{grad3} using $\bbB_2$
        \ENDFOR
        \STATE Compute the forecasting $\text{NMSE}_p$ for this window
        \IF{$\frac{\text{NMSE}_p}{\text{NMSE}_{\min}} - 1 < \epsilon$}
            \STATE Retain cell $\bbe(p) \gets 1$, update $\text{NMSE}_{\min} = \text{NMSE}_p$
        \ELSE
            \STATE Restore cached states $(\bbx, \bbP)$ and $(\bbalpha, \bbh, \bbGamma)$
            \STATE Revert $\bbe(p) \gets 0$, update $\bbB_2 \gets \bbB_2^* \mathrm{diag}(\bbe)$
            \FOR{$i = T_s + (p{-}1) \frac{T{-}T_s}{N_2}$ \TO $T_s + p \frac{T{-}T_s}{N_2}$}
                \STATE Re-run updates via~\eqref{grad1}--\eqref{grad3} using corrected $\bbB_2$
            \ENDFOR
        \ENDIF
    \ENDFOR
    \RETURN $\bbB_2$
\end{algorithmic}
}
\end{algorithm}

\begin{figure}[t]
\centering
\resizebox{\linewidth}{!}{%
\begin{tikzpicture}[
    scale=1.0,
    time_tick/.style={rectangle, fill=black, inner sep=0pt, minimum height=10pt, minimum width=2pt},
    brace/.style={decorate, decoration={brace, amplitude=8pt}, gray!80, thick},
    flow_box/.style={rectangle, draw=blue!40, fill=white, rounded corners, font=\small, align=center, thick, inner sep=4pt}
]

\draw[-{Latex}, thick] (-0.5, 0) -- (11.5, 0) node[right, font=\normalsize, text=black] {Time steps};

\fill[gray!20] (0,-0.5) rectangle (3,0.5);
\node[font=\small, text=black, align=center] at (1.5,0) {Warm-up\\[-0.5ex]($T_s$ steps)};

\fill[blue!15] (3,-0.5) rectangle (4.5,0.5); \node[font=\small, text=blue!80!black] at (3.75,0) {Cell 1};
\fill[blue!10] (4.5,-0.5) rectangle (6,0.5); \node[font=\small, text=blue!80!black] at (5.25,0) {Cell 2};

\node at (7.25, 0) [text=gray] {...}; 

\fill[blue!15] (8.5,-0.5) rectangle (10,0.5); \node[font=\small, text=blue!80!black] at (9.25,0) {Cell $N_2$};

\foreach \x in {0, 3, 4.5, 6, 8.5, 10} {
    \draw[thick, white] (\x,-0.5) -- (\x,0.5); 
    \node[time_tick] at (\x, 0) {};
}

\node[below=3pt, font=\normalsize] at (0,-0.5) {$T_0=0$};
\node[below=3pt, font=\normalsize, fill=white, inner sep=1.5pt, rounded corners=2pt] at (3,-0.5) {$T_s$};
\node[below=3pt, font=\normalsize] at (10,-0.5) {$T$};

\draw[brace] (3, 0.7) -- node[above=8pt, font=\bfseries\normalsize, text=orange] {Identification Phase} (10, 0.7);

\draw[densely dashed, gray, thick] (3, -0.5) -- (-0.2, -1.2);
\draw[densely dashed, gray, thick] (4.5, -0.5) -- (11.8, -1.2);

\filldraw[fill=blue!5, draw=blue!40, thick, rounded corners] (-0.2, -4.5) rectangle (11.8, -1.2);
\node[font=\bfseries\normalsize, text=blue!80!black] at (5.6, -1.6) {Detail: Inside Window $p$ (Snapshot \& Rollback)};

\node[flow_box] (step1) at (1.4, -2.4) {\textbf{1. Snapshot}\\$\bbe(p) \gets 1$};
\node[flow_box] (step2) at (4.2, -2.4) {\textbf{2. Window EKF}\\Compute $\text{NMSE}_p$};
\node[flow_box, draw=orange!50, fill=orange!5] (step3) at (7.0, -2.4) {\textbf{3. Evaluate}\\Ratio $< \epsilon$?};

\node[flow_box, draw=green!60!black, fill=green!5] (accept) at (9.6, -2.4) {\textbf{Accept}\\Commit state};

\node[flow_box, draw=red!60, fill=red!5] (reject) at (7.0, -3.7) {\textbf{Reject \& Rollback}\\Reject the cell \& Re-run\\this window};

\draw[-{Latex}, thick, gray!80] (step1) -- (step2);
\draw[-{Latex}, thick, gray!80] (step2) -- (step3);

\draw[-{Latex}, thick, green!60!black] (step3.east) -- node[midway, above, font=\small] {Yes} (accept.west);
\draw[-{Latex}, thick, red] (step3.south) -- node[midway, right, font=\small] {No} (reject.north);

\draw[densely dashed, gray] (accept.east) -| (11.3, -3.0);
\draw[densely dashed, gray] (reject.east) -| (11.3, -3.0);
\draw[-{Latex}, densely dashed, gray] (11.3, -3.0) -- (12.2, -3.0) node[right, font=\scriptsize, align=center] {Next\\window};

\end{tikzpicture}
}
\caption{Online $2$-cell identification algorithm. Following a $T_s$-step warm-up, candidate $2$-cells are evaluated across $N_2$ discrete windows. In each window, the EKF state is first cached and a candidate $2$-cell is activated. If the forecasting $\mathrm{NMSE}_p$ reduction meets the threshold $\epsilon$, the $2$-cell is accepted. If rejected, the system executes a rollback to the cached state and re-runs the EKF using the original topology present at the start of the window.}
\label{fig:cell_id}
\end{figure}
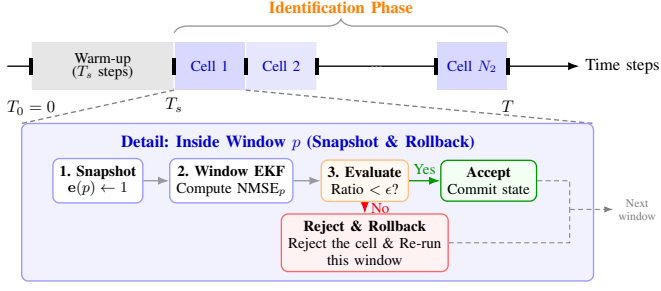
The complete procedure is illustrated in Figure~\ref{fig:cell_id} and detailed in Algorithm~\ref{alg:online_cell}. Initially, we consider a warm-up phase of $T_s$ iterations without $2$-cells. This allows the filter to stabilize, yielding an average error, $\mathrm{NMSE}_0$, which serves as the initial benchmark $\mathrm{NMSE}_{\min} = \mathrm{NMSE}_0$. 

During the identification phase, we evaluate the candidate $2$-cells in descending order of their uncertainty scores $\xi_p$. For the $p$-th selected candidate, we repeat the following steps:
\begin{itemize}[wide, labelindent=0pt, labelsep=0.5em, nosep]
    \item \textbf{Snapshot and Evaluate:} We cache the current EKF states and parameters. We then activate the candidate $2$-cell ($\bbe(p)\leftarrow 1$), update the incidence matrix $\bbB_2 \gets \bbB_2^* \mathrm{diag}(\bbe)$, and execute the parameter updates \eqref{grad1}--\eqref{grad3} over the designated time window $t \in [T_s + (p{-}1) \frac{T{-}T_s}{N_2}, T_s + p \frac{T{-}T_s}{N_2}]$ to compute the trial forecasting NMSE $\mathrm{NMSE}_p$.
    \item \textbf{Accept or Rollback:} If the relative error ratio satisfies the tolerance threshold $\epsilon$, i.e., $\frac{\mathrm{NMSE}_p}{\mathrm{NMSE}_{\min}} - 1 < \epsilon$, the trial is deemed successful. We retain the $2$-cell and update the tracking benchmark: $\mathrm{NMSE}_{\min} = \mathrm{NMSE}_p$. Conversely, if the threshold is not met, the candidate $2$-cell degrades predictive performance. In this case, we restore the cached states and parameters, revert $\bbe(p)\leftarrow 0$, and re-run the EKF updates over the time window using the input topology of this window.
\end{itemize}

The finite horizon $T$ in Algorithm~\ref{alg:online_cell} defines an initial topology identification phase. In an infinite horizon setting ($T \to \infty$), the framework can be extended to operate continuously. Instead of terminating after evaluating $N_2$ candidates, the system can periodically infer the structure.

\vspace{-0.3cm}
\section{Frequency Analysis}\label{S:Frequency}

\subsection{Topological Subspaces and Spectral Representations}

\subsubsection{Hodge Decomposition}
The algebraic structure of the Hodge Laplacian induces an orthogonal decomposition of the $k$-order signal space $\mathbb{R}^{N_k}$ into three mutually orthogonal subspaces: the gradient $\operatorname{im}(\mathbf{B}_k^\top)$, the curl $\operatorname{im}(\mathbf{B}_{k+1})$, and the harmonic $\operatorname{ker}(\mathbf{L}_k)$. Consequently, $k$-signal $\boldsymbol{s}^k$ decomposes as:
\begin{equation}\label{eq:hodge_signal_decomp}
\boldsymbol{s}^k = \boldsymbol{s}_{\mathrm{G}}^k + \boldsymbol{s}_{\mathrm{H}}^k + \boldsymbol{s}_{\mathrm{C}}^k = \mathbf{B}_k^\top \underline{\boldsymbol{s}}^{k-1} + \boldsymbol{s}_\mathrm{H}^k + \mathbf{B}_{k+1} \overline{\boldsymbol{s}}^{k+1},
\end{equation}
where $\underline{\boldsymbol{s}}^{k-1}$ and $\overline{\boldsymbol{s}}^{k+1}$ denote lower and higher-order potentials, respectively, and $\mathbf{L}_k \boldsymbol{s}_\mathrm{H}^k = \mathbf{0}$. For instance, edge signals ($1$-signals) partition into irrotational gradient flows, solenoidal curl flows, and globally conservative harmonic flows.

This spatial decomposition is intrinsically linked to the eigendecomposition of the Hodge Laplacian, $\mathbf{L}_k = \mathbf{U}_k \boldsymbol{\Lambda}_k \mathbf{U}_k^\top$. By organizing the orthonormal basis as $\mathbf{U}_k = [\mathbf{U}_{k,\mathrm{G}} \; \mathbf{U}_{k,\mathrm{C}} \; \mathbf{U}_{k,\mathrm{H}}]$---spanning the gradient, curl, and harmonic subspaces---we define the Topological Fourier Transform (TFT) as $\hat{\boldsymbol{s}}^k = \mathbf{U}_k^{\top} \boldsymbol{s}^k$ \cite{barbarossa2020topological}. Unlike graphs, this cellular spectrum natively isolates distinct physical behaviors, projecting the signal into independent gradient ($\hat{\boldsymbol{s}}^k_{\mathrm{G}}$), curl ($\hat{\boldsymbol{s}}^k_{\mathrm{C}}$), and harmonic ($\hat{\boldsymbol{s}}^k_{\mathrm{H}}$) spectral coefficients.


\vspace{-0.3cm}
\subsection{Process Dynamics}
By collecting the intra-order spectral components, we define the block-diagonal matrices $\mathbf{U} = \mathrm{blkdiag}(\mathbf{U}_0, \mathbf{U}_1, \mathbf{U}_2)$ and $\boldsymbol{\Lambda} = \mathrm{blkdiag}(\boldsymbol{\Lambda}_0, \boldsymbol{\Lambda}_1, \boldsymbol{\Lambda}_2)$. Here, $\mathbf{U}$ serves as the complete cell Fourier basis that jointly diagonalizes the global block Laplacian $\mathbf{L} = \mathrm{blkdiag}(\mathbf{L}_0, \mathbf{L}_1, \mathbf{L}_2)$. Applying the Fourier transform $\hat{\boldsymbol{x}} = \mathbf{U}^\top \boldsymbol{x}$ to the state equation, and substituting the eigendecomposition $\mathbf{L} = \mathbf{U} \boldsymbol{\Lambda} \mathbf{U}^\top$, diagonalizes the deterministic process dynamics:

{\small{
\begin{equation}\label{eq:spectral_dynamics}
\hat{\boldsymbol{x}}_i = (\mathbf{I} - c\,\delta_t\,\boldsymbol{\Lambda})\,\hat{\boldsymbol{x}}_{i-1} + \hat{\boldsymbol{q}}_i, 
\end{equation}}}

where the spectral process noise $\hat{\boldsymbol{q}}_i \sim \mathcal{N}(\mathbf{0}, \tilde{\mathbf{Q}})$ has its covariance matrix explicitly defined by:{\small{
\begin{equation}\label{eq:spectral_Q}
\tilde{\mathbf{Q}} \triangleq \mathbf{U}^\top \mathbf{Q} \mathbf{U} \simeq \delta_t\, \mathbf{U}^\top \mathbf{D}\, \mathrm{diag}^2(\boldsymbol{\alpha})\, \mathbf{D}^\top \mathbf{U}.
\end{equation}}}
To make this explicit, the scalar evolution of the $k$-th topological frequency component is given as:
\begin{equation}\label{eq:scalar_spectral}
\hat{x}_i[k] = (1 - c\,\delta_t\,\lambda_k)\hat{x}_{i-1}[k] + \hat{q}_i[k].
\end{equation}

Existing topological time-series models~\cite{krishnan2024simplicial,toplor2025} primarily rely on the Hodge Laplacians $\bbL$ to describe signal evolution. Although effective, these approaches are fundamentally limited by the eigenvalue multiplicity problem. Due to the presence of structural symmetries in real-world networks, the Hodge Laplacian frequently yields multiple identical eigenvalues. Consequently, these features receive identical filter responses, rendering them spectrally indistinguishable. To demonstrate how the proposed formulation resolves this eigenvalue multiplicity, we analyze the spectral process noise covariance $\tilde{\bbQ}$. Consider two orthonormal eigenmodes, $\bbu_w$ and $\bbu_r$, sharing a degenerate non-zero eigenvalue $\lambda_w = \lambda_r = \lambda > 0$. Defining the boundary flows as $\bbv_w = \bbD^\top \bbu_w$ and $\bbv_r = \bbD^\top \bbu_r$, the marginal stochastic variance for mode $w$ expands as:

\vspace{-0.5cm}
{\small{\begin{equation}\label{eq:expanded_variance}
\tilde{\bbQ}_{w,w} = \delta_t \bbu_w^\top \bbD \mathrm{diag}^2(\bbalpha) \bbD^\top \bbu_w = \delta_t \sum_{j=1}^{N} \alpha_j^2 [\bbv_w]_j^2.
\end{equation}}}

To establish the necessity of a spatially varying $\bbalpha$, consider the counter-example of a uniform uncertainty model where $\alpha_j = c$ for all $j$. Under this assumption, the constant scalar $c^2$ factors out of the summation, and the variance algebraically collapses back into the Laplacian eigenvalue:
\begin{equation}
\tilde{\bbQ}_{w,w} = \delta_t c^2 (\bbu_w^\top \bbD \bbD^\top \bbu_w) = \delta_t c^2 (\bbu_w^\top \bbL \bbu_w) = \delta_t c^2 \lambda.
\end{equation}
Because mode $r$ shares the exact same eigenvalue $\lambda$, it trivially yields an identical variance $\tilde{\bbQ}_{r,r} = \delta_t c^2 \lambda$, failing to break the spectral degeneracy. However, in our framework, the spatially heterogeneous uncertainty vector $\bbalpha$ naturally breaks this degeneracy by decoupling the variances based on their local spatial support. Let $\mathbf{p}_w$ and $\mathbf{p}_r$ denote the spatial energy distributions, where $[\mathbf{p}_w]_j = [\bbv_w]_j^2$ and $[\mathbf{p}_r]_j = [\bbv_r]_j^2$. In topology, degenerate modes typically arise from identical but physically separated substructures (e.g., disjoint cyclic loops). Consequently, $\mathbf{p}_w$ and $\mathbf{p}_r$ occupy different subsets of the network. Because the inner products $\tilde{\bbQ}_{w,w} = \delta_t \mathbf{p}_w^\top \bbalpha^{2}$ and $\tilde{\bbQ}_{r,r} = \delta_t \mathbf{p}_r^\top \bbalpha^{2}$ act as spatial masks, the variance of mode $w$ is determined entirely by the localized uncertainty parameters in its specific region, while mode $r$ is determined by an entirely different regional subset of $\bbalpha^{2}$. Since real-world networks exhibit localized, non-uniform physical characteristics (e.g., localized traffic variations or regional sensor noise), these disjoint parameter subsets are inherently unequal. This physical decoupling effectively breaks the symmetry and ensures $\tilde{\bbQ}_{w,w} \neq \tilde{\bbQ}_{r,r}$. While different marginal process variances do not strictly guarantee independent tracking in isolation, they crucially provides the EKF  the necessary priors to better separate spectrally overlapping modes.

{\vspace{-.45cm}\subsection{ Observation Analysis}\label{S:Observability}
Having established the process dynamics in the frequency domain, we next analyze the conditions under which each frequency mode is observable from the observations. Note that for clarity of exposition we provide the analysis for linear model.  Recalling $\boldsymbol{y}_i^k$ denote the observation vector at the $k$-th topological order at time step $i$. Under a spatial sampling mask $\mathbf{\Phi}_k$, the spatial observation matrix corresponding to the $k$-cell observations is formulated as:}
{\small{\begin{equation}
    \mathbf{C}_k = \mathbf{\Phi}_k \begin{bmatrix} \mathbf{B}_k^\top \mathbf{H}_{k, k-1}(\mathbf{L}_{k-1}) & \mathbf{H}_{k,k}(\mathbf{L}_k) & \mathbf{B}_{k+1} \mathbf{H}_{k, k+1}(\mathbf{L}_{k+1}) \end{bmatrix}.
\end{equation}}}

\noindent where $\boldsymbol{y}_i^k=\bbC_k\bbx_i$.

First we projecting the spatial observations into the frequency domain, $\hat{\boldsymbol{y}}_i^k = \mathbf{U}_k^\top \boldsymbol{y}_i^k$. Assuming full observability momentarily ($\mathbf{\Phi}_k = \mathbf{I}$), the corresponding fully spectral observation matrix is formed as $\hat{\mathbf{C}}_k = \mathbf{U}_k^\top \mathbf{C}_k \mathbf{U}$. 

Because the topological polynomial filters commute with the Fourier bases, yielding $\mathbf{H}(\mathbf{L}_j)\mathbf{U}_j = \mathbf{U}_j \hat{\mathbf{H}}(\mathbf{\Lambda}_j)$, and leveraging the orthogonality $\mathbf{U}_k^\top \mathbf{U}_k = \mathbf{I}$, $\hat{\mathbf{C}}_k$ evaluates explicitly to $\hat{\mathbf{C}}_k = \begin{bmatrix} \hat{\mathbf{C}}_{k,1} & \hat{\mathbf{C}}_{k,2} & \hat{\mathbf{C}}_{k,3} \end{bmatrix}$, where the constituent spectral blocks are defined as:{\small{
\begin{align}
    \hat{\mathbf{C}}_{k,1} &= \mathbf{U}_k^\top \mathbf{B}_k^\top \mathbf{U}_{k-1} \hat{\mathbf{H}}_{k, k-1}(\mathbf{\Lambda}_{k-1}), \\
    \hat{\mathbf{C}}_{k,2} &= \hat{\mathbf{H}}_{k,k}(\mathbf{\Lambda}_k), \\
    \hat{\mathbf{C}}_{k,3} &= \mathbf{U}_k^\top \mathbf{B}_{k+1} \mathbf{U}_{k+1} \hat{\mathbf{H}}_{k, k+1}(\mathbf{\Lambda}_{k+1}).
\end{align}}}
The structural observability of the entire cross-order system is dictated by the rank of the fully spectral observability matrix $\hat{\boldsymbol{\mathcal{O}}} = [\hat{\mathbf{C}}_k^\top, (\hat{\mathbf{C}}_k\tilde{\mathbf{\Lambda}})^\top, \ldots, (\hat{\mathbf{C}}_k\tilde{\mathbf{\Lambda}}^{N-1})^\top]^\top$.

To determine the rank of $\hat{\boldsymbol{\mathcal{O}}}$, we analyze the subspace projections induced by the generalized boundary operators. Based on the Hodge decomposition, any $k$-order signal space partitions into mutually orthogonal subspaces: $\mathbb{R}^{N_k} = \operatorname{im}(\mathbf{B}_k^\top) \oplus \operatorname{im}(\mathbf{B}_{k+1}) \oplus \operatorname{ker}(\mathbf{L}_k)$. By arranging the $k$-th order TFT basis according to this decomposition, $\mathbf{U}_k = [\mathbf{U}_{k,G} \;\; \mathbf{U}_{k,C} \;\; \mathbf{U}_{k,H}]$, and letting $\mathcal{Q}_G, \mathcal{Q}_C, \mathcal{Q}_H$ denote the respective frequency bands.
Specifically, lower-order signals act as a gradient-like embedding $\bbx_{i,G}^k \triangleq \mathbf{B}_k^\top \mathbf{H}_{k, k-1}(\mathbf{L}_{k-1})\bbx_i^{k-1}$, while higher-order signals inject a curl-like embedding $\bbx_{i,C}^k \triangleq \mathbf{B}_{k+1} \mathbf{H}_{k, k+1}(\mathbf{L}_{k+1})\bbx_i^{k+1}$. Because the fundamental topological identity dictates $\mathbf{B}_k\mathbf{B}_{k+1} = \mathbf{0}$, the image spaces of these cross-order mappings are orthogonal, satisfying $\operatorname{im}(\mathbf{B}_k^\top) \perp \operatorname{im}(\mathbf{B}_{k+1})$.

Consequently, projecting the spatial embeddings into the spectral domain via the $k$-th order TFT yields sparse representations. By defining the spectral embeddings as $\hat{\bbx}_{i,G}^k \triangleq \mathbf{U}_k^\top \bbx_{i,G}^k$ and $\hat{\bbx}_{i,C}^k \triangleq \mathbf{U}_k^\top \bbx_{i,C}^k$, the full spectral observation equation decomposes elegantly as:

{\small{
\begin{equation}
    \hat{\boldsymbol{y}}_i^k = \hat{\mathbf{H}}_{k,k}(\mathbf{\Lambda}_k)\hat{\bbx}_i^k + \hat{\bbx}_{i,G}^k + \hat{\bbx}_{i,C}^k + \hat{\mathbf{n}}_i^k.
\end{equation}}}

 These cross-order embedding vectors are the images of the adjacent states mapped through the outer blocks of the spectral observation matrix: $\hat{\bbx}_{i,G}^k = \hat{\mathbf{C}}_{k,1}\hat{\bbx}_i^{k-1}$ and $\hat{\bbx}_{i,C}^k = \hat{\mathbf{C}}_{k,3}\hat{\bbx}_i^{k+1}$. Because the fundamental identity $\mathbf{B}_k\mathbf{B}_{k+1} = \mathbf{0}$ ensures the column spaces of $\hat{\mathbf{C}}_{k,1}$ and $\hat{\mathbf{C}}_{k,3}$ are structurally orthogonal, the generated vectors $\hat{\bbx}_{i,G}^k$ and $\hat{\bbx}_{i,C}^k$ are rigidly confined to disjoint frequency supports ($\mathcal{Q}_G$ and $\mathcal{Q}_C$). Under full sampling ($\mathbf{\Phi}_k = \mathbf{I}$), this geometric orthogonality guarantees that the cross-order terms contribute $r_{k,G} + r_{k,C}$ linearly independent dimensions to $\operatorname{rank}(\hat{\mathbf{C}}_k)$, where $r_{k,G} = \operatorname{rank}(\mathbf{B}_k)$ and $r_{k,C} = \operatorname{rank}(\mathbf{B}_{k+1})$. Because $\hat{\mathbf{C}}_k$ forms the foundational block row of $\hat{\boldsymbol{\mathcal{O}}}$, this expanded column space directly increases the rank of the entire spectral observability matrix. 

It is important to clarify that the structural decoupling induced by boundary-map orthogonality does not strictly equate to formal full observability of the LTI system. From a control-theoretic perspective, strict observability would further require satisfying algebraic rank conditions (e.g., the PBH test) involving $\hat{\mathbf{C}}_k$, ensuring non-zero filter responses across all modes, and explicitly resolving the spectral degeneracy associated with repeated eigenvalues of $\tilde{\mathbf{\Lambda}}$. Therefore, the analysis presented herein serves as an interpretive structural framework rather than a strict observability guarantee.
{\vspace{-.3cm}\subsection{Special Case: Observability of the 1-Skeleton and 2-Cells}
\subsubsection{Node and Face Modes via Edge Observations}
When only edge signals $\boldsymbol{y}_i^1$ are observed, the adjacent latent states can be inferred due to the spectral decoupling:}
\begin{itemize}[wide, labelindent=0pt, labelsep=0.5em, nosep]
    \item \textbf{Node Modes:} The node embedding $\hat{\bbx}_{i,G}^1$ is mathematically injected exclusively into the gradient frequency band $\mathcal{Q}_G$. Because it is completely decoupled from the curl and harmonic modes, the relative node dynamics are observable by decoding the $\mathcal{Q}_G$ spectrum of the edge signals, provided the topological filter response $\hat{\mathbf{H}}_{10}(\mathbf{\Lambda}_0)$ is non-degenerate. Algebraically, this guarantees that the associated block column of the observation matrix contributes $r_G = \operatorname{rank}(\mathbf{B}_1)$ linearly independent dimensions to $\operatorname{rank}(\hat{\mathbf{C}}_1)$.
    \item \textbf{Face Modes:} Similarly, the rotational face embedding $\hat{\bbx}_{i,C}^1$ projects exclusively into the curl band $\mathcal{Q}_C$. The fundamental boundary identity ensures complete decoupling from the gradient modes. Consequently, the latent face modes can be recovered by isolating the $\mathcal{Q}_C$ frequency band of the edge observations. Correspondingly, this orthogonal cross-order mapping injects $r_C = \operatorname{rank}(\mathbf{B}_2)$ independent dimensions into the rank of the observation matrix $\hat{\mathbf{C}}_1$.
\end{itemize}

\subsubsection{Reconstruction of Edge Modes from Node or Face Observations}
The discrete spectrum of the boundary operators symmetrically facilitates the recovery of latent edge signals from node or face observations.
\begin{itemize}[wide, labelindent=0pt, labelsep=0.5em, nosep]
    \item \textbf{Edge Gradient Modes from Nodes:} If only node signals $\boldsymbol{y}_i^0$ are measured, the observation relies on the divergence mapping $\mathbf{B}_1 \bbx_i^1$. In the spectral domain, this projection takes the form $\hat{\boldsymbol{y}}_i^0 = \mathbf{U}_0^\top \mathbf{B}_1 \bbx_i^1 = \hat{\mathbf{B}}_1 \hat{\bbx}_{i,G}^1$. Because $\hat{\mathbf{B}}_1$ exhibits full column rank over the non-zero spectrum, the corresponding observation matrix preserves $r_G$ independent dimensions, ensuring the irrotational edge flows ($\hat{\bbx}_{i,G}^1$) are entirely observable from the measured node spectra.
    \item \textbf{Edge Curl Modes from Faces:} Conversely, if only face signals $\boldsymbol{y}_i^2$ are observed, the observation relies on the curl mapping $\mathbf{B}_2^\top \bbx_i^1$. The spectral projection yields $\hat{\boldsymbol{y}}_i^2 = \mathbf{U}_2^\top \mathbf{B}_2^\top \bbx_i^1 = \hat{\mathbf{B}}_2^\top \hat{\bbx}_{i,C}^1$. This exact mapping guarantees the observation matrix retains $r_C$ linearly independent dimensions, permitting the reconstruction of the solenoidal edge flows ($\hat{\bbx}_{i,C}^1$) directly from the measured face spectra.
\end{itemize}

It is crucial to explicitly note that adjacent-order observations cannot recover the harmonic components of the edge signals. Because harmonic flows $\hat{\bbx}_{i,H}^1$ belong to the kernel of the Hodge Laplacian, they are strictly divergence-free ($\mathbf{B}_1 \bbx_{i,H}^1 = \mathbf{0}$) and curl-free ($\mathbf{B}_2^\top \bbx_{i,H}^1 = \mathbf{0}$).

\vspace{-0.15cm}
\section{Numerical experiments}\label{S:exp}
We benchmark our model against graph and topological baselines on two tasks: time-series forecasting (with complete and incomplete data) and topology inference (recovering the underlying cell complex).
{\vspace{-.6cm}\subsection{Datasets}}
\subsubsection{Synthetic}
The first dataset is generated to match our model's assumptions, serving to corroborate our theoretical framework and demonstrate that the algorithm performs as expected in controlled conditions. The constructed topology consists of 10 nodes, 17 edges, and 9 cells. The latent state model is parameterized with $c=0.5$, $\delta t=0.1$, and process noise variance $\sigma_p=1$, while the remaining hyperparameters are randomly initialized. The observation model is $\boldsymbol{y}_i = 10 \bbH(\bbL)\cos(\bbx_i) + \bbn_i$, and the observation noise variance is $\sigma_o=1$. Both latent states and observations are generated with a sequence length of 1000 time steps.

\subsubsection{Cherry Hills}
The second dataset involves a real-world water distribution network modeled as a cell complex: nodes represent storage tanks, junctions or reservoirs, edges are pipes, and higher-order cells are closed pipe loops. The network includes 36 nodes, 40 edges, and 5 cells. We consider node pressure levels and edge flow rates, generated via the EPANET simulator under a demand-driven model~\cite{krishnan2024simplicial, liu2025matched}. The resulting dataset comprises 661 time steps, spanning 55 hours of measurements sampled every five minutes.


\subsubsection{Wireless Sensor Network}
The third dataset features wireless sensor readings from the Intel Berkeley Research Lab~\cite{intel2004}. We model the network as a cell complex: nodes are sensors, edges are defined based on pairwise distance thresholds, and cells represent enclosed regions. We retain the 15 most informative sensors with the most complete data records, resulting in a topology with 15 nodes, 26 edges, and 12 cells. Signals include node-level humidity and edge-level voltage differences. This setup captures how humidity affects electrical impedance and voltage drops~\cite{chen2005humidity}. The dataset consists of 1000 time steps.

\subsubsection{Capital Bikeshare Network}
The fourth dataset uses Capital Bikeshare trip data~\cite{capitalbikeshare} from the Washington, D.C. area. We model the 10 most active stations as nodes, the 15 busiest routes as edges, and the resulting seven loops as cells. Data is aggregated daily: node signals represent total station visits, and edge signals reflect bidirectional traffic volume. Using trip records from January 2023 to October 2025, the final dataset contains 1034 time steps.
{\vspace{-.5cm}\subsection{Forecasting}
We consider the forecasting task and evaluate performance using the normalized mean squared error $\operatorname{NMSE}(t)=\frac{1}{N} \sum_{i=1}^N \frac{\sum_{\tau=1}^t\left(y_i(\tau)-\hat{y}_i(\tau)\right)^2}{\sum_{\tau=1}^t y_i^2(\tau)}$,
where $N$ denotes the dimension of the signal components, $y_i(\tau)$ represents the ground truth value of the $i-$th component at time step $\tau$, and $\hat{y}_i(\tau)$ denotes the corresponding forecasting value. The models are trained online to predict the observations at the subsequent time step. An initial portion of the data is reserved for validation, hyperparameter selection and parameter stabilization, with the percentage $p_{\text{stab}}$ varying across datasets to account for their specific characteristics.

\subsubsection{Models}
We forecast the next signal given historical observations, assuming full observability ($\boldsymbol{\Phi} = \mathbf{I}$). Hyperparameters are grid-searched. We report the average normalized mean squared error (NMSE) and standard deviation over five independent runs for the following models:
\begin{itemize}[wide, labelindent=0pt, labelsep=0.5em, nosep]
    \item \textbf{TKF-N/L (Dirac)}: Our proposed nonlinear (Algorithm~\ref{alg:TKF}) and linear models. Parameters are summarized in Table~\ref{tab:parameters}. TKF-L isolates the impact of nonlinearity.
    \item \textbf{TKF-N/L (Hodge)}: Ablated variants using Hodge Laplacians, removing cross-order topological coupling.
    \item \textbf{TKF (empty)}: A variant ignoring 2-cells. Unlike G-KNet~\cite{sabbaqi2025gknet}, it simultaneously processes node and edge signals. Filter order is 3 ; other parameters match TKF-N.
    \item \textbf{TOPO-LMS}~\cite{toplor2025}: An LMS-based adaptive filter for simplicial complexes. Filter order is 1 (searched $\in \{1,2,3\}$).
    \item \textbf{SC-VAR}~\cite{krishnan2024simplicial} \& \textbf{CC-VAR}~\cite{canbolat2025cellular}: Vector autoregressive models for simplicial and cell complexes, respectively.
    \item \textbf{LSTM}~\cite{hochreiter1997long}: An offline-trained RNN to highlight the need for topological inductive biases. Hidden size is 256 (searched $\in [32, 1024]$) and window size is 5 (searched $\in [1, 20]$).
\end{itemize}

\begin{table}[t]
\centering
\caption{Parameters used by TKF-N for each experiment}
\footnotesize{
\begin{tabular}{l|lllllll}
   \toprule
   Datasets  & $M$ & $\sigma_k$& $\sigma_o$& $p_{\text{stab}}$& $T$  & $c$ & $\delta t$\\
   \midrule
   Synthetic & 2   & 5         & 1         & 0.1              & 1000 & 0.5 & 0.1\\
   Cherry    & 2   & 5         & 1         & 0.01             & 661  & 0.5 & 0.1\\
   WSN       & 10  & 5         & 1         & 0.1              & 1000 & 0.5 & 0.1\\
   Capital Bikeshare   & 2   & 5         & 1         & 0.01             & 1034 & 0.5 & 0.1\\
   \bottomrule
\end{tabular}
}
\label{tab:parameters}
\end{table}

\subsubsection{Results}
\begin{table}[t]
\centering
\caption{NMSE of the forecasting with complete data.}
\label{tab:forecasting}
\resizebox{\columnwidth}{!}{%
\begin{tabular}{l|cccc}
   \toprule
   \textbf{Method} & \textbf{Synthetic} & \textbf{Cherry} & \textbf{WSN} & \textbf{Capital Bikeshare}\\
   \midrule
   TKF-N (Dirac)   & \textbf{0.018$\pm$0.008}  & 0.083$\pm$0.004  & \textbf{0.014$\pm$0.002} & 0.220$\pm$0.006 \\
   TKF-L (Dirac)   & 0.018$\pm$0.005  & 0.086$\pm$0.005  & 0.016$\pm$0.002 & 0.218$\pm$0.001 \\
   \midrule
    TKF-N (Hodge)   & 0.536$\pm$0.052  & 0.069$\pm$0.008  & 0.042$\pm$0.001 & 0.296$\pm$0.006 \\
    TKF-L (Hodge)   & 0.536$\pm$0.053  & 0.067$\pm$0.009  & 0.041$\pm$0.001 & 0.297$\pm$0.004 \\
   \midrule
   TKF (empty)      & 0.080$\pm$0.039  & 0.086$\pm$0.004  & 0.014$\pm$0.002 & \textbf{0.210$\pm$0.004} \\
   \midrule
   TOPO-LMS ~\cite{toplor2025}& 0.065$\pm$0.033  & 0.031$\pm$0.001  & 0.017$\pm$0.010 & 0.270$\pm$0.001 \\
   SC-VAR ~\cite{krishnan2024simplicial}      & 0.535$\pm$0.067  & 0.149$\pm$0.079  & 0.039$\pm$0.004 & 0.307$\pm$0.008 \\
   CC-VAR ~\cite{canbolat2025cellular}      & 0.044$\pm$0.002  & 0.085$\pm$0.002  & 0.122$\pm$0.010 & 0.705$\pm$0.006 \\
   LSTM ~\cite{hochreiter1997long}& 0.121$\pm$0.005  & \textbf{0.030$\pm$0.002}  & 0.022$\pm$0.005 & 0.222$\pm$0.006 \\
   \bottomrule
\end{tabular}%
}
\end{table}

\begin{figure*}[htbp]
    \centering
    \hspace{-1cm}
    \includegraphics[width=0.9\textwidth]{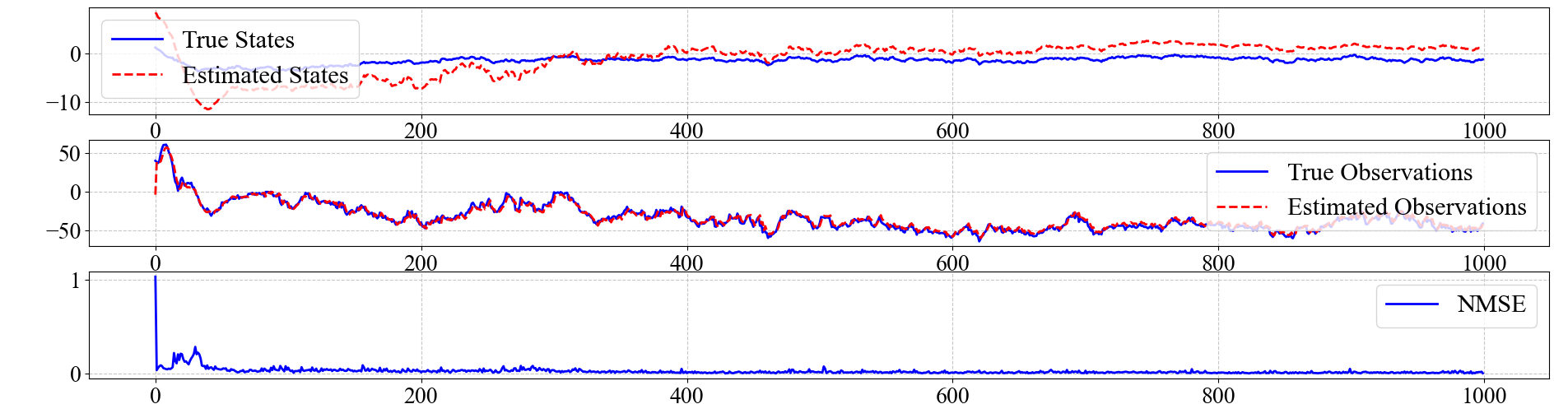}
    \caption{One-step-ahead forecasting on synthetic data. The top row shows the true latent states (blue solid lines) versus estimated states (red dashed lines). The middle row compares true observations (blue solid lines) with model forecasting (red dashed lines). The bottom row shows the evolution of the NMSE.}
    \label{fig:forecasting}
\end{figure*}
Table~\ref{tab:forecasting} shows the forecasting results. Overall, the proposed TKF achieves the best performance because it explicitly models the underlying system dynamics. TKF-N generally outperforms its linear counterpart across most datasets. An exception is the Capital Bikeshare dataset, where TKF-L performs slightly better, likely due to less dominant nonlinear relationships. The advantage of TKF-N is twofold: first, its nonlinear assumption maps latent states to observations in real-world conditions; second, the learnable nonlinear coefficients $\bbGamma$ alongside the convolution filter coefficients $\bbh$ increases model expressivity.

TKF (empty)'s performance is inferior because it ignores higher-order information. TOPO-LMS assumes wide-sense stationarity, which fails for non-stationary real-world signals, preventing it from fully capturing temporal dynamics. Similarly, SC-VAR and CC-VAR are fundamentally limited by their linear nature and lack of a latent state-space model. Relying solely on direct linear autoregression without a latent "true state," they struggle to separate complex underlying dynamics from measurement noise. Finally, although LSTM captures nonlinear dynamics, its black-box nature ignores the underlying topological geometry and higher-order connectivity.

Our ablation study confirms that joint cross-order processing (Dirac) generally outperforms intra-order filtering (Hodge) which validates our theoretical intuition: working independently per order discards vital topological dependencies.

In Fig.~\ref{fig:forecasting}, we present the tracking curve for the synthetic dataset, the only dataset for which ground-truth states are available. Remarkably, the TKF can accurately track the observations and also partially track the latent state. Although the loss function does not explicitly include constraints on the latent state, the model can follow its dynamics, indicating that the proposed framework captures the underlying system behavior.
{\vspace{-.3cm}\subsection{Forecasting with Missing Data}
We now evaluate the robustness of the proposed model by performing  forecasting with missing data. This experiment  demonstrate that TKF can maintain high predictive accuracy even when portions of the input history are unavailable. TOPO-LMS was not included in the comparison because it cannot perform autoregressive forecasting for missing values.}

\subsubsection{Experimental setup}
We randomly removed 10\%, 20\%, and 30\% of the signal entries in each dataset. The experiment was repeated five times, and the results are reported in terms of the mean and standard deviation of the NMSE.

\subsubsection{Results}

\begin{table}[t]
\centering
\caption{Performance (NMSE) of forecasting under partial observability across different datasets.}
\label{tab:interpolation}
\footnotesize{
\begin{tabular}{l|ccc}
\toprule
\multirow{2}{*}{Method} & \multicolumn{3}{c}{Synthetic} \\
 & 10\% & 20\% & 30\% \\
\midrule
TKF-N (Dirac) & \textbf{0.051$\pm$0.026} & \textbf{0.108$\pm$0.079} & \textbf{0.135$\pm$0.077} \\
TKF-L (Dirac) & 0.168$\pm$0.183 & 0.195$\pm$0.107 & 0.386$\pm$0.311 \\
TKF-N (Hodge) & 0.562$\pm$0.054 & 0.656$\pm$0.114 & 0.651$\pm$0.084 \\
TKF-L (Hodge) & 0.569$\pm$0.058 & 0.653$\pm$0.114 & 0.624$\pm$0.089 \\
TKF (empty)    & 0.247$\pm$0.088 & 1.028$\pm$1.196 & 1.174$\pm$0.994 \\
\midrule
\multirow{2}{*}{Method} & \multicolumn{3}{c}{Cherry} \\
 & 10\% & 20\% & 30\% \\
\midrule
TKF-N (Dirac) & 0.175$\pm$0.098 & 0.506$\pm$0.197 & 0.628$\pm$0.281 \\
TKF-L (Dirac) & 0.169$\pm$0.096 & 0.479$\pm$0.178 & 0.595$\pm$0.265 \\
TKF-N (Hodge) & \textbf{0.118$\pm$0.054} & \textbf{0.198$\pm$0.119} & 0.349$\pm$0.263 \\
TKF-L (Hodge) & 0.136$\pm$0.059 & 0.207$\pm$0.105 & \textbf{0.309$\pm$0.182} \\
TKF (empty)    & 0.178$\pm$0.100 & 0.499$\pm$0.199 & 0.638$\pm$0.296 \\
\midrule
\multirow{2}{*}{Method} & \multicolumn{3}{c}{WSN} \\
 & 10\% & 20\% & 30\% \\
\midrule
TKF-N (Dirac) & 0.120$\pm$0.038 & 0.218$\pm$0.079 & 0.360$\pm$0.091 \\
TKF-L (Dirac) & 0.115$\pm$0.021 & \textbf{0.194$\pm$0.085} & \textbf{0.250$\pm$0.062} \\
TKF-N (Hodge) & \textbf{0.099$\pm$0.037} & 1.153$\pm$2.082 & 4.835$\pm$7.548 \\
TKF-L (Hodge) & 0.164$\pm$0.080 & 0.258$\pm$0.120 & 3.747$\pm$5.399 \\
TKF (empty)    & 0.103$\pm$0.035 & 0.201$\pm$0.068 & 0.321$\pm$0.083 \\
\midrule
\multirow{2}{*}{Method} & \multicolumn{3}{c}{Capital Bikeshare} \\
 & 10\% & 20\% & 30\% \\
\midrule
TKF-N (Dirac) & 0.903$\pm$0.323 & 1.167$\pm$0.325 & 1.489$\pm$0.242 \\
TKF-L (Dirac) & 0.999$\pm$0.588 & 1.120$\pm$0.378 & 1.399$\pm$0.293 \\
TKF-N (Hodge) & 0.329$\pm$0.037 & 0.796$\pm$0.558 & 0.659$\pm$0.328 \\
TKF-L (Hodge) & \textbf{0.318$\pm$0.012} & \textbf{0.707$\pm$0.456} & \textbf{0.584$\pm$0.396} \\
TKF (empty)    & 0.688$\pm$0.176 & 0.782$\pm$0.163 & 1.290$\pm$0.399 \\
\bottomrule
\end{tabular}%
}
\end{table}

Table~\ref{tab:interpolation} summarizes the results. While NMSE for all methods increases with data missingness, our model remains resilient. Their robustness stems from using higher-order interactions to constrain signal reconstruction.

\begin{figure*}[htbp]
    \centering
    \hspace{-1cm}
    \includegraphics[width=0.9\textwidth]{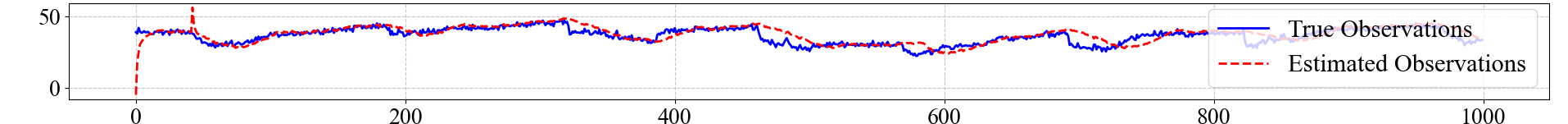}

    \hspace{-1cm}
    \includegraphics[width=0.9\textwidth]{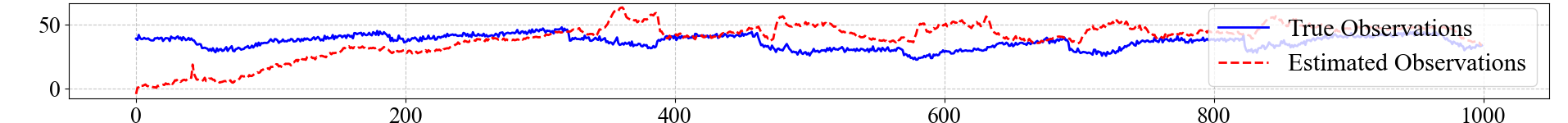}
    \caption{Forecasting comparison between complete and incomplete signals on Wireless Sensor Network dataset. The subplot above shows the observation tracking performance with complete signal while the subplot below shows the incomplete signal with 30\% of the missing values. The observations tracking curve is plotted for an unobserved signal component.}
    \label{fig:interpolation}
\end{figure*}

Interestingly, the TKF-L outperforms TKF-N in some incomplete data cases. Under partial observability, latent state reconstruction becomes an ill-posed inverse problem. In these data-scarce regimes, the highly flexible nonlinear mapping in TKF-N overfits to the limited available signals, amplifying errors during forecasting. Conversely, the linear mapping in TKF-L imposes a strict, lower-order structural prior which acts as a natural regularizer, making TKF-L more robust.

As the missing rate increases, TKF(empty) is worse compared to the variants consider the higher-order structures, demonstrating that the information provided by higher-order structures significantly enhances the model's robustness.

Figure~\ref{fig:interpolation} shows tracking performance on the Wireless Sensor Network (WSN) dataset under varying missing ratios. This dataset is highlighted as it exhibits the most significant performance gaps between models. As missing data increases, deviations between predicted and actual observations grow, and NMSE convergence slows, requiring more iterations to stabilize. Furthermore, steady-state NMSE remains higher than in the fully observed case, reflecting the challenge of forecasting with incomplete information.
{\vspace{-.45cm}\subsection{Cell Identification}
We compare the proposed cell identification algorithm with the topology inference approach implemented in TOPO-LMS. The performance of both methods is evaluated in terms of inference accuracy, precision, recall and F1 score. We conduct experiments on synthetic data where the ground-truth topology is known, allowing for a quantitative assessment of the proposed algorithm against the baseline. We also compared the NMSEs corresponding to empty, inferred, and full topological structures in table ~\ref{tab:nmse} to show the effectiveness of our method.}

\subsubsection{Experimental setup}
We set $T_s = 0.1T$ to stabilize the training parameters for initialization. The NMSE change threshold is varied from $0$ to $-0.1$. We vary the proportion of cells included in the ground-truth topology of the synthetic data from $0$ to $1.0$. Each experiment is repeated $10$ times independently, and the average results are reported.

\subsubsection{Results}
\begin{figure}[t]
    \centering
    \includegraphics[width=0.5\textwidth]{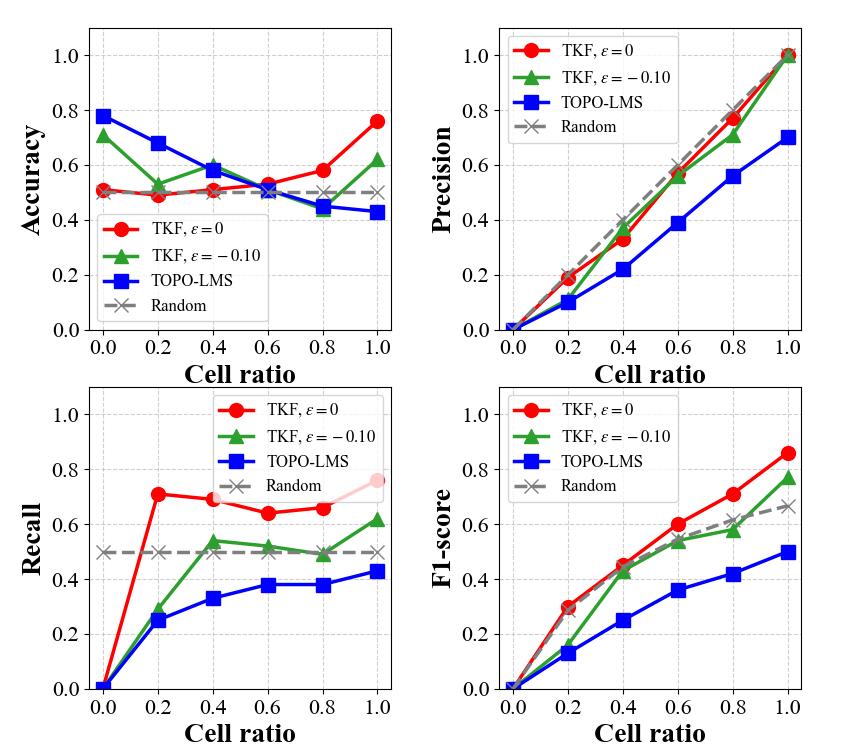}
    \caption{ 
The proposed TKF algorithm is compared with TOPO-LMS across varying cell ratios in terms of four evaluation metrics: Accuracy, Precision, Recall, and F1-score.}
    \label{fig:identification}
\end{figure}

Fig.~\ref{fig:identification} compares the performance of the proposed TKF algorithm, TOPO-LMS, and a random predictor across varying $2$-cell ratios. At a sparse $2$-cell ratio, the signal lacks sufficient information about higher-order interactions, preventing the model from effectively inferring the  higher-order structure. A smaller tolerance parameter $\epsilon$ makes the model more conservative which is evidenced at $\epsilon=-0.10$, where the algorithm yields a lower recall and precision compared with $\epsilon=0$. This indicates that in environments with high structural uncertainty, the proposed algorithm prioritizes the suppression of false positives over the detection of rare true positives. As the $2$-cell ratio increases, the increasing topological information mitigates this problem. For $\epsilon=0$, the recall significantly increases to 0.71 at a cell ratio of 0.2, surpassing the 0.5 random baseline. The F1-score of TKF exhibits a monotonic increase, peaking at 0.86 with an accuracy of 0.76 at a cell ratio of 1.0. The stricter variant with $\epsilon=-0.10$ follows a similar trend, reaching a peak F1-score of 0.77, which reflects a trade-off between rigorous acceptance criteria and detection rates. The baseline TOPO-LMS fails to effectively infer the underlying topology. This limitation stems fundamentally from its reliance on a linear signal model.

Table~\ref{tab:nmse} evaluates the forecasting NMSE under empty, inferred, and full topological structures. For the synthetic dataset, the inferred topology consistently outperforms the empty baseline. The predictive accuracy of our algorithm matches that of the full cell baseline. This shows that the additional network cells are redundant and provide no extra predictive value. By achieving the same performance with a smaller subset, our approach proves it can extract a minimal yet completely sufficient representation of the network. For the real-world datasets, the NMSE values across most datasets are slightly better except for the Capital Bikeshare. This indicates that the algorithm maintains robustness to NMSE forecasting accuracy while inferring the structure.

\begin{table}[t]
\centering
\caption{Performance (NMSE) of forecasting for the inferred structure.}
\label{tab:nmse}
\resizebox{\columnwidth}{!}{%
\begin{tabular}{ll|ccc}
\toprule
\multirow{2}{*}{Dataset} & \multirow{2}{*}{Percentage} & \multicolumn{3}{c}{Observability} \\
\cmidrule{3-5}
 & & Empty & Inferred & Full \\
\midrule
\multirow{6}{*}{Synthetic} 
 & 0\%   & 0.044$\pm$0.020 & 0.042$\pm$0.020 & 0.041$\pm$0.020 \\
 & 20\%  & 0.052$\pm$0.025 & 0.038$\pm$0.016 & 0.035$\pm$0.017 \\
 & 40\%  & 0.076$\pm$0.038 & 0.036$\pm$0.014 & 0.031$\pm$0.012 \\
 & 60\%  & 0.081$\pm$0.033 & 0.034$\pm$0.015 & 0.027$\pm$0.012 \\
 & 80\%  & 0.083$\pm$0.047 & 0.034$\pm$0.016 & 0.023$\pm$0.007 \\
 & 100\% & 0.080$\pm$0.039 & 0.026$\pm$0.013 & 0.018$\pm$0.008 \\
\midrule
Cherry  & Unknown & 0.086$\pm$0.004 & 0.085$\pm$0.004 & 0.083$\pm$0.004 \\
WSN     & Unknown & 0.014$\pm$0.002 & 0.014$\pm$0.002 & 0.014$\pm$0.002 \\
Capital Bikeshare & Unknown & 0.210$\pm$0.004 & 0.216$\pm$0.004 & 0.220$\pm$0.006 \\
\bottomrule
\end{tabular}%
}
\end{table}
{\vspace{-.45cm}\section{Conclusion} \label{S:conclusion}
In this paper, we proposed a topology-aware state space framework to infer latent dynamics and capture higher-order interactions from multivariate time-series defined over cell complexes. Grounded in stochastic partial differential equations (SPDEs), our approach models state evolution as a  topological diffusion governed by Hodge Laplacians and Dirac operators. To handle partial observability and measurement nonlinearities, we introduced a cellular convolution based observation function. For recursive state estimation, we employed an Extended Kalman  with point-wise nonlinearity. Filter, coupled with an online Expectation-Maximization algorithm to learn model parameters and state uncertainties. Furthermore, to address scenarios with incomplete structural knowledge, we introduced a heuristic online cell identification mechanism that  infers missing higher-order topologies directly from data. The experiments on synthetic and real-world datasets confirm that our approach yields reliable estimates and recovers underlying structures. }
\vspace{-.45cm}
\bibliographystyle{IEEEtran}
\bibliography{mybib_final}
\end{document}